\documentstyle[preprint,aps,epsfig]{revtex}
\tightenlines
\begin{document}
\draft
\preprint{\begin{minipage}[b]{1.5in}
          UK/TP 99-19\\
          \end{minipage}}
\vspace{0.2in}

\title{Nuclear dependence coefficient $\alpha(A,q_T)$ for the Drell-Yan 
       and J/$\psi$ production}
\author{Xiaofeng Guo$^1$, Jianwei Qiu$^2$, and Xiaofei Zhang$^{2}$}
\address{$^1$Department of Physics and Astronomy,
             University of Kentucky,\\
             Lexington, Kentucky 40506, USA\\
         $^2$Department of Physics and Astronomy,
             Iowa State University \\
             Ames, Iowa 50011, USA }
\date{April 10, 2000}
\maketitle
\begin{abstract}
Define the nuclear dependence coefficient $\alpha(A,q_T)$ in terms of 
ratio of transverse momentum spectrum in hadron-nucleus and in
hadron-nucleon collisions: $\frac{d\sigma^{hA}}{dq_T^2}/
\frac{d\sigma^{hN}}{dq_T^2}\equiv A^{\alpha(A,q_T)}$.  We argue that
in small $q_T$ region, the $\alpha(A,q_T)$ for the Drell-Yan and
J/$\psi$ production is given by a universal function:\ $a+b\, q_T^2$,
where parameters $a$ and $b$ are completely determined by either
calculable quantities or independently measurable physical
observables.  We demonstrate that this universal function
$\alpha(A,q_T)$ is insensitive to the $A$ for normal nuclear targets.
For a color deconfined nuclear medium, the $\alpha(A,q_T)$ becomes
strongly dependent on the $A$.  We also show that our $\alpha(A,q_T)$ 
for the Drell-Yan process is naturally linked to perturbatively calculated
$\alpha(A,q_T)$ at large $q_T$ without any free parameters, and the
$\alpha(A,q_T)$ is consistent with E772 data for all $q_T$. 
\end{abstract}
\vspace{0.2in}

\section{Introduction}
\label{sec:1}

It has long been observed that in high energy hadron-nucleus
collisions, the transverse momentum spectra of produced particles
differ significantly from that in hadron-nucleon collisions
\cite{Cronin,Cronin-data}.  This anomalous nuclear dependence is
known as the Cronin effect. In recent years, many data on such nuclear 
dependence for the Drell-Yan \cite{NA10,E772,Eloss} and J/$\psi$
production~\cite{E866,Jpsi-Et,NA3} became available, and these new
data have renewed our interests to understand the observed novel
effect \cite{alpha-thy}.  Although it is believed that the Cronin effect
is a result of parton multiple scattering inside nuclear medium, 
systematic calculations in QCD for such anomalous nuclear dependence
only exist for the large transverse momentum ($q_T$) region
\cite{LQS1,Guo1,BMuller}, but not for the small $q_T$ region due to 
technical difficulties in handling multi-scale calculations in QCD 
perturbation theory.  In this paper, we derive the nuclear dependence 
in transverse momentum distributions for the Drell-Yan and 
J/$\psi$ production in the small $q_T$ region by
combining constraints from perturbatively calculable quantities in
the large $q_T$ region and the available information from independently
measured physical observables \cite{GQZ1}. 

In high energy nuclear collisions, in addition to short-distance
single scattering, parton multiple scattering becomes very important.
According to QCD factorization theorem \cite{CSS-fac}, physical
observables due to single hard scattering in hadronic collisions
depend on the parton distributions.  In terms of generalized QCD
factorization theorem \cite{QS-fac}, contributions to physical
observables from parton multiple scattering are directly proportional
to multiparton correlation functions.  These correlation functions are 
as fundamental as the parton distributions, and they provide
complementary information on nonperturbative QCD dynamics \cite{LQS2}.
Since single hard scattering is localized in position space, it is the 
multiple scattering that is sensitive to the size of nuclear medium,
and is therefore responsible for the anomalous nuclear dependence.
Measurements of such anomalous nuclear dependence provide a window of
opportunities to explore the dynamics of parton correlations, which
have important implications for  the physics in relativistic
heavy ion collisions.  

The nuclear dependence of the transverse momentum spectrum 
is often presented in terms of the ratio of cross sections $R$ or the 
nuclear dependence coefficient $\alpha$.  For example, for the Drell-Yan
pair production, the $R$ and $\alpha$ are defined by   
\begin{equation}
R(A,q_T)\equiv  \frac{1}{A}\, \left. 
                \frac{d\sigma^{hA}}{dQ^2 dq_T^2} \right/
\frac{d\sigma^{hN}}{dQ^2 dq_T^2} \, 
\equiv \, A^{\alpha(A,q_T)-1} 
\label{R-qt}
\end{equation}
where $Q$ and $q_T$ are the Drell-Yan pair's total invariant mass and
transverse momentum, respectively, and $A$ is the atomic weight
of the nuclear target. In Eq.~(\ref{R-qt}), 
$d\sigma^{hA}/dQ^2 dq_T^2$ and 
$d\sigma^{hN}/dQ^2 dq_T^2$ are the transverse momentum spectrum
in hadron-nucleus and hadron-nucleon collisions, respectively. 
Since single hard scattering is localized in space, both $R(A,q_T)$
and $\alpha(A,q_T)$ should be very close to one if there is no
multiple scattering.  
However, data on the ratio of the Drell-Yan transverse
momentum spectrum show that $R(A,q_T)$ very much differs from one (or
$\alpha(A,q_T)$ is significant away from one) \cite{E772}.  In
Ref.~\cite{E772}, data on $R(A,q_T)$ shows a nontrivial dependence on 
both $A$ and $q_T$ for different nuclear targets (including C, Ca, Fe,
and W).  In terms of $\alpha(A,q_T)$, the Drell-Yan data has the 
following general features: $\alpha(A,q_T=0)$ is less than one, it
increases as $q_T$ increases, and it can be as large as 1.07
\cite{NA10,E772,Eloss}.   
     
When $q_T$ is large ($\sim Q$), the $q_T$-dependence of $\alpha(A,q_T)$
can be calculated within QCD perturbation theory, and a significant
nuclear enhancement in $\alpha(A,q_T)$ was predicted \cite{Guo1}.
Although it was argued in Ref.~\cite{Guo1} that in the small $q_T$
region, the $\alpha(A,q_T)$ or the transverse momentum spectrum should 
show nuclear suppression, no quantitative analysis or prediction was
given for the suppression.  Since almost all existing data on the 
nuclear dependence of the Drell-Yan production are in the small $q_T$
region \cite{NA10,E772,Eloss}, it is very important to derive a
quantitative description of the nuclear dependence coefficient
$\alpha(A,q_T)$ or the ratio $R(A,q_T)$ for the small $q_T$ region.

Furthermore, J/$\psi$ data in hadron-nucleus collisions show that the
$\alpha(A,q_T)$ (or $R(A,q_T)$) for J/$\psi$ production as a function
of $q_T$ has similar features as that of the $\alpha(A,q_T)$ (or
$R(A,q_T)$) for the Drell-Yan production \cite{Jpsi-Et,NA3}.  Recent
data from Fermilab experiment E866 show that $\alpha(A,q_T)$ for
J/$\psi$ production has an universal shape, but its magnitude depends
on the range of $x_F$ \cite{E866}. J/$\psi$ suppression in
relativistic heavy ion collisions was predicted to signal the color
deconfinement \cite{Satz}.  Recently, significant J/$\psi$ suppression
has been observed in existing fixed target experiments
\cite{JPsi-sup,NA50}.  
The data have generated a lot of theoretical discussions 
in searching for the mechanism of 
the observed J/$\psi$ suppression.  Understanding the
features in $\alpha(A,q_T)_{{\rm J/}\psi}$ is very valuable for 
such investigations. 
 
In this paper, we argue that in the small $q_T$ region, $\alpha(A,q_T)$
for the Drell-Yan production is given by an universal function:\
$a_1+b_1\,q_T^2$, where $a_1$ and $b_1$ are completely determined by 
the measured ratio of total cross sections, 
$\frac{d\sigma^{hA}}{dQ^2}/\frac{d\sigma^{hN}}{dQ^2}$, and the averaged  
transverse momentum square $\langle q_T^2 \rangle^{hN}_{DY}$ in
hadron-nucleon collisions, plus perturbatively calculable quantities,
such as the transverse momentum broadening 
$\Delta\langle q_T^2\rangle_{DY}$ in hadron-nucleus collisions.
The $\langle q_T^2\rangle^{hN}_{DY}$ and $\Delta\langle
q_T^2\rangle_{DY}$ will be defined in Eqs.~(\ref{qt2}) and 
(\ref{dydqt2}).

According to the generalized QCD factorization theorem \cite{QS-fac},
like all perturbatively calculable hadronic quantities, the transverse
momentum broadening for the Drell-Yan production, $\Delta\langle
q_T^2\rangle_{DY}$, can be factorized into a convolution of an
infrared safe hard part and corresponding universal quark-gluon
correlation functions.  The infrared safe hard part is calculable in
QCD perturbation theory and was derived in Ref.~\cite{Guo2}. 
Although the quark-gluon correlation functions are nonperturbative in
nature and unknown, just like the well-known parton distributions,
these functions are universal and they appear in the factorized
expressions of other physical observables.  For example, the
same quark-gluon correlation functions in the expression for 
$\Delta\langle q_T^2\rangle_{DY}$ appear in the factorized formulas
for the $\alpha(A,q_T)$ in the large $q_T$ region \cite{Guo1}. 
Because of the universality of these quark-gluon correlation functions, 
the data on $\Delta\langle q_T^2\rangle_{DY}$ can be used  
to extract the quark-gluon correlation functions, which can then be used 
to predict $\alpha(A,q_T)$ in the large $q_T$ region.  In this paper, 
we show that the Fermilab E772 data on $\Delta\langle
q_T^2\rangle_{DY}$ can be used to extract the size of the quark-gluon
correlation functions, which can then be used to predict the
$\alpha(A,q_T)$ in the small $q_T$ region.

We further demonstrate that our $\alpha(A,q_T)$ for the Drell-Yan
production in small $q_T$ region is naturally connected to the
perturbatively calculated $\alpha(A,q_T)$ in the large $q_T$ region,
and we also show that the predicted $\alpha(A,q_T)$ is consistent with
the E772 data in both the small and large $q_T$ regions.  Furthermore,
we show that although the ratio of the transverse momentum spectrum
$R(A,q_T)$ can  
have a non-trivial dependence on the atomic weight $A$, the nuclear
dependence coefficient $\alpha(A,q_T)$ is {\it insensitive} to the
atomic weight $A$ for normal nuclear targets.  On the other hand, we
argue that in a color deconfined nuclear medium, the $\alpha(A,q_T)$
becomes very sensitive to the $A$ due to a long range color
correlation.  

In addition, we show that $\alpha(A,q_T)$ for J/$\psi$ production 
in small $q_T$ region is
given by the same functional form:\ $a_2+b_2\, q_T^2$, and the new
parameters $a_2$ and $b_2$ are also completely determined by the
the ratio of total cross sections
$\sigma^{hA}_{{\rm J/}\psi}/\sigma^{hN}_{{\rm J/}\psi}$ and the averaged
transverse momentum square $\langle q_T^2 \rangle^{hN}_{{\rm J/}\psi}$
in hadron-nucleon collisions, plus perturbatively calculable
quantities, such as the transverse momentum broadening
$\Delta\langle q_T^2\rangle_{{\rm J/}\psi}$ in hadron-nucleus
collisions. Our predictions for $\alpha_{{\rm J/}\psi}(A,q_T)$ are
consistent with recent data from Fermilab E866.   

The predicting power of our nuclear dependence coefficient function 
$\alpha(A,q_T)$ for the Drell-Yan and J/$\psi$ production 
is its universal quadratic dependence on $q_T$, and the
fact that all parameters are completely fixed by either calculable or
independently measurable quantities.  In addition, we predict that the
$\alpha(A,q_T)$ is extremely insensitive to the atomic number 
$A$ for normal nuclear 
targets while it can be very sensitive to A (or the medium size) for a
color deconfined medium.

The rest of this paper is organized as follows.  In the next section,
we argue that the Drell-Yan transverse momentum spectrum in
hadron-nucleon and hadron-nucleus collisions should be well
represented by a Gaussian-like distribution in the small $q_T$ region
at fixed target energies.  From such a transverse momentum spectrum,
we derive a universal expression of $\alpha(A,q_T)$ for the Drell-Yan 
production in the small $q_T$ region in Sec.~\ref{sec:3}.  We
explicitly demonstrate that the  
$\alpha(A,q_T)$ is insensitive to the atomic weight $A$, and   
the $\alpha(A,q_T)$ in
the small $q_T$ region is naturally connected to the perturbatively
calculated $\alpha(A,q_T)$ in the large $q_T$ region. 
We then compare our universal function $\alpha(A,q_T)$ 
with E772 data at all $q_T$.  
In Sec.~\ref{sec:4}, we provide $\alpha(A,q_T)$ for J/$\psi$
production, and discuss the similarity and difference between the
$\alpha(A,q_T)$ for the Drell-Yan and J/$\psi$ production.  
we show that in a color deconfined nuclear medium, the $\alpha(A,q_T)$
for J/$\psi$ production becomes very sensitive to the atomic number 
$A$ (or the medium size), while it is
extremely insensitive to the $A$ for normal color confined nuclear
targets.  Finally, in Sec.~\ref{sec:5}, we summarize our main
conclusions, and  discuss the predicting power of our
nuclear dependence coefficient $\alpha(A,q_T)$.

\section{The Drell-Yan transverse momentum spectrum}
\label{sec:2}

\begin{figure}
\epsfig{figure=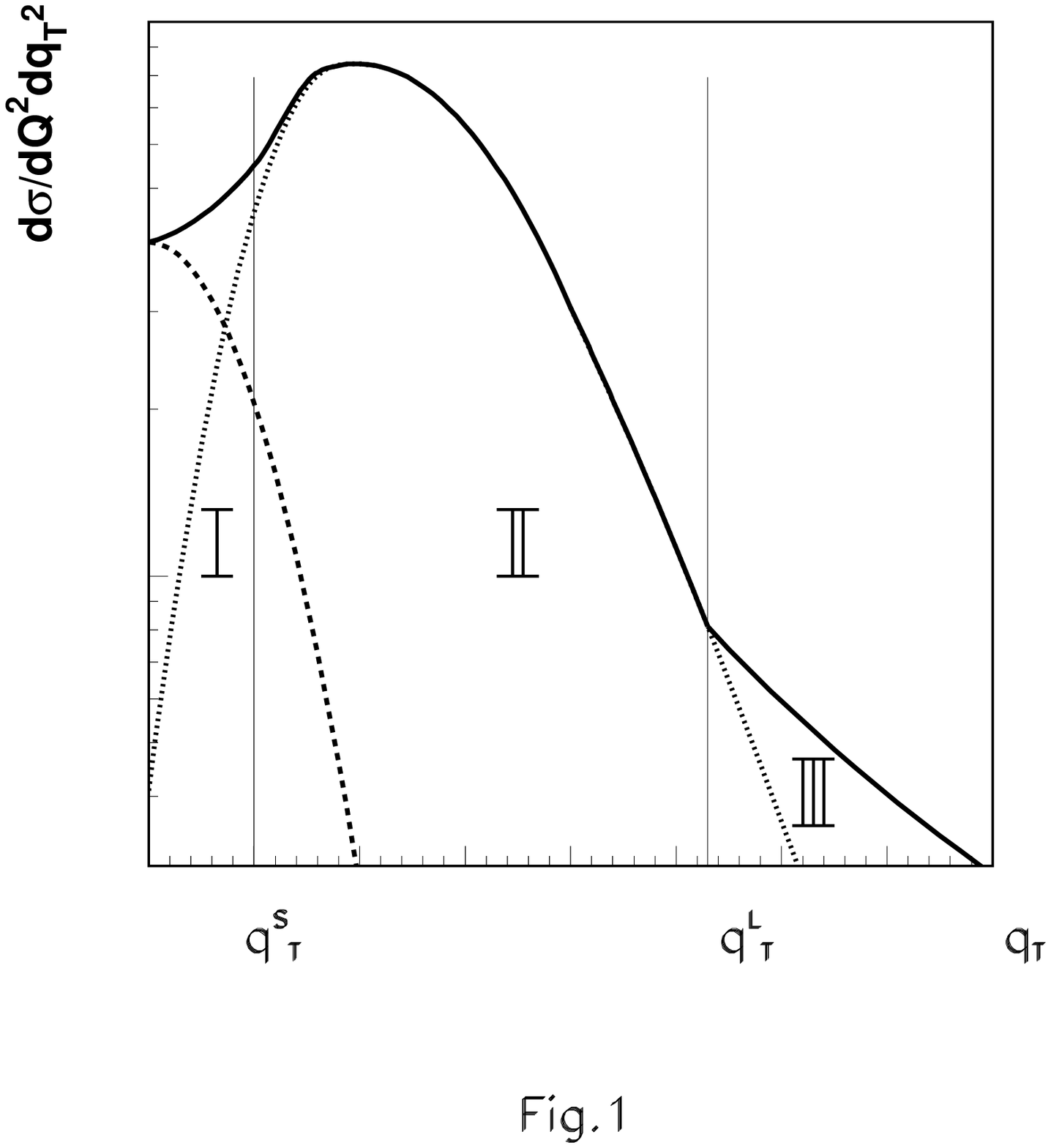,width=1.1in}
\caption{Typical Drell-Yan transverse momentum spectrum as a function
of $q_T$ in three regions. }
\label{fig1}
\end{figure}

Depending on physics origins, the Drell-Yan transverse momentum spectrum
can be divided into three regions, as shown in Fig.~\ref{fig1}.  The
small $q_T$ region corresponds to the region where $q_T< q_T^S \sim 1$~GeV.  
The spectrum in this region, labeled by I in Fig.~\ref{fig1}, is dominated
by the intrinsic transverse momenta of colliding partons.  At leading
order in perturbation theory, the Drell-Yan transverse momentum
distribution is given by 
\begin{equation}
\frac{d\sigma^{hN}}{dQ^2 d^2q_T} = \frac{d\sigma_{hN}}{dQ^2} \,
\delta^2(\vec{q}_T)\ .
\label{DYLO}
\end{equation}
The effect of the parton intrinsic transverse momentum can be included by
replacing the $\delta$-function in Eq.~(\ref{DYLO}) by its
Gaussian-representation, $\delta(q_x)={\rm lim}_{\tau\rightarrow 0}\,
\frac{1}{\sqrt{2\pi}\,\tau}\, 
{\rm exp}[-q_x^2/2\tau^2]$ \cite{ESW}, and we obtain
\begin{equation}
\frac{d\sigma^{(I)}}{dQ^2 dq_T^2} = N_{DY}\, \frac{1}{2\tau^2}\,
{\rm e}^{-q_T^2/2\tau^2}\, , 
\label{DYGau}
\end{equation}
where the superscript ``(I)'' indicates the region ``I'', $N_{DY}$ is
a dimensional normalization, and $\tau$ is the 
width of the Gaussian-like distribution.  The physical meaning of the
$N_{DY}$ and $\tau$ will be determined later. In principle, 
$N_{DY}$ and $\tau$ are non-perturbative quantities.  

The intermediate $q_T$ region, labeled by II in Fig.~\ref{fig1},
corresponds to the region where $q_T^S \leq q_T \leq q_T^L$ with 
$q_T^L\equiv \kappa Q$ and $\kappa \sim 1/3 - 1/2$.  In this region,
the physically observed scales: $Q$ and $q_T$ are both large enough,
and in principle, QCD perturbation theory can be used to calculate the
short-distance partonic scattering and the Drell-Yan cross section
can be factorized into the following form \cite{CSS-fac}
\begin{equation}
\frac{d\sigma^{hN}}{dQ^2\,dq_T^2} = \sum_{a,b}\,
\phi_{a/h}\left(x',\mu^2\right)\otimes
\phi_{b/N}\left(x,\mu^2\right)\otimes 
\frac{d\hat{\sigma}^{ab}}{dQ^2\,dq_T^2}
\left(x',x,\alpha_s(\mu^2),\frac{Q^2}{\mu^2},\frac{q_T^2}{\mu^2}\right)
\label{DY-fact}
\end{equation}
where $\sum_{a,b}$ sum over all parton flavors $a$ and $b$, the $\mu$
represents both the renormalization and factorization scales, the
$\phi$'s and $\otimes$'s represent the parton distributions and the
convolution over incoming partons' momentum fractions, respectively. 
In Eq.~(\ref{DY-fact}), the $d\hat{\sigma}^{ab}/dQ^2\,dq_T^2$ is a
short-distance contribution for partons $a$ and $b$ to produce a
Drell-Yan lepton pair, and it can be calculated perturbatively in
terms of a power series of the strong coupling constant
$\alpha_s(\mu^2)$ with the coefficients that depend logarithmically on
the ratio of scales: $Q^2/\mu^2$ and $q_T^2/\mu^2$.  The scale $\mu$
can be chosen to be equal to $Q$ or $q_T$, or somewhere between, and
the factorized Drell-Yan cross section in Eq.~(\ref{DY-fact}) should
not be sensitive to a specific choice of the scale if the factorized
expression is reliable. 

If the $Q$ is much larger than the $q_T$, the perturbatively
calculated $d\hat{\sigma}^{ab}/dQ^2\,dq_T^2$ will have large
logarithms, such as $\ln(Q^2/q_T^2)$, for any choice of the scale
$\mu$.  These large logarithms are due to the gluon radiation off the
incoming partons before the hard collision takes place to produce
a Drell-Yan lepton pair \cite{Resum,CSS-resum}.  Because of the
infrared and collinear singularities associated with the massless
gluons, we can have two powers of the large logarithms for every power
of the strong coupling constant $\alpha_s$ in the perturbatively
calculated $d\hat{\sigma}^{ab}/dQ^2\,dq_T^2$.  Clearly, the
logarithms, $(\alpha_s\ln^2(Q^2/q_T^2))^n$, can be larger than the
unity if the $Q$ is much larger than the $q_T$.  For example, for
$Z^0$ production at the Tevatron energies, the ratio $Q^2/q_T^2$ can
be as large as $8\times 10^{3}$ for $q_T\sim 1$~GeV and $Q\sim
91$~GeV.  Consequently, the high order corrections in $\alpha_s$ can
be as important as the lower order terms, and therefore, the
perturbatively calculated $d\hat{\sigma}^{ab}/dQ^2\,dq_T^2$ at a fixed
order of $\alpha_s^n$ are no longer reliable.  

Using the renormalization group method, the large logarithms,
$(\alpha_s\ln^2(Q^2/q_T^2))^n$, in the perturbatively calculated 
$d\hat{\sigma}^{ab}/dQ^2\,dq_T^2$ can be resummed to
all orders in $\alpha_s$ \cite{Resum,CSS-resum}.  As pointed out in
Ref.~\cite{CSS-resum}, the resummation of the large logarithms can be 
systematically carried out by solving a corresponding renormalization
group equation in the $b$-space, which is the Fourier
transform of the transverse momentum $q_T$-space.  The kernel of
the renormalization group equation can be calculated order by order in
$\alpha_s$ in QCD perturbation theory.  In order to obtain the
Drell-Yan transverse momentum distribution, one has to perform the
Fourier transform of the $b$-space solution of the renormalization
group equation 
back to the $q_T$ space; and it is necessary to have the $b$-space
solution for all values of $b$.  However, since the kernel was
calculated in perturbative QCD, the $b$-space solution is only
reliable for small values of $b$ (i.e., $b\ll 1/\Lambda_{QCD}$).
Therefore, a nonperturbative function $F^{NP}(b)$ had to be introduced
to cover the large $b$ region \cite{CSS-resum}.  The parameters in the
$F^{NP}(b)$ were determined by fitting experimental data on the Drell-Yan
$q_T$ spectrum \cite{Davies,Yuan}.  The role of this nonperturbative
function $F^{NP}(b)$ is similar to that of the input parton
distributions $\phi(x,Q_0^2)$ when we solve the renormalization group
equations (or the DGLAP equations) for the parton distributions' 
$Q^2$-dependence.  

\begin{figure}
\begin{minipage}[t]{2.0in}
\epsfig{figure=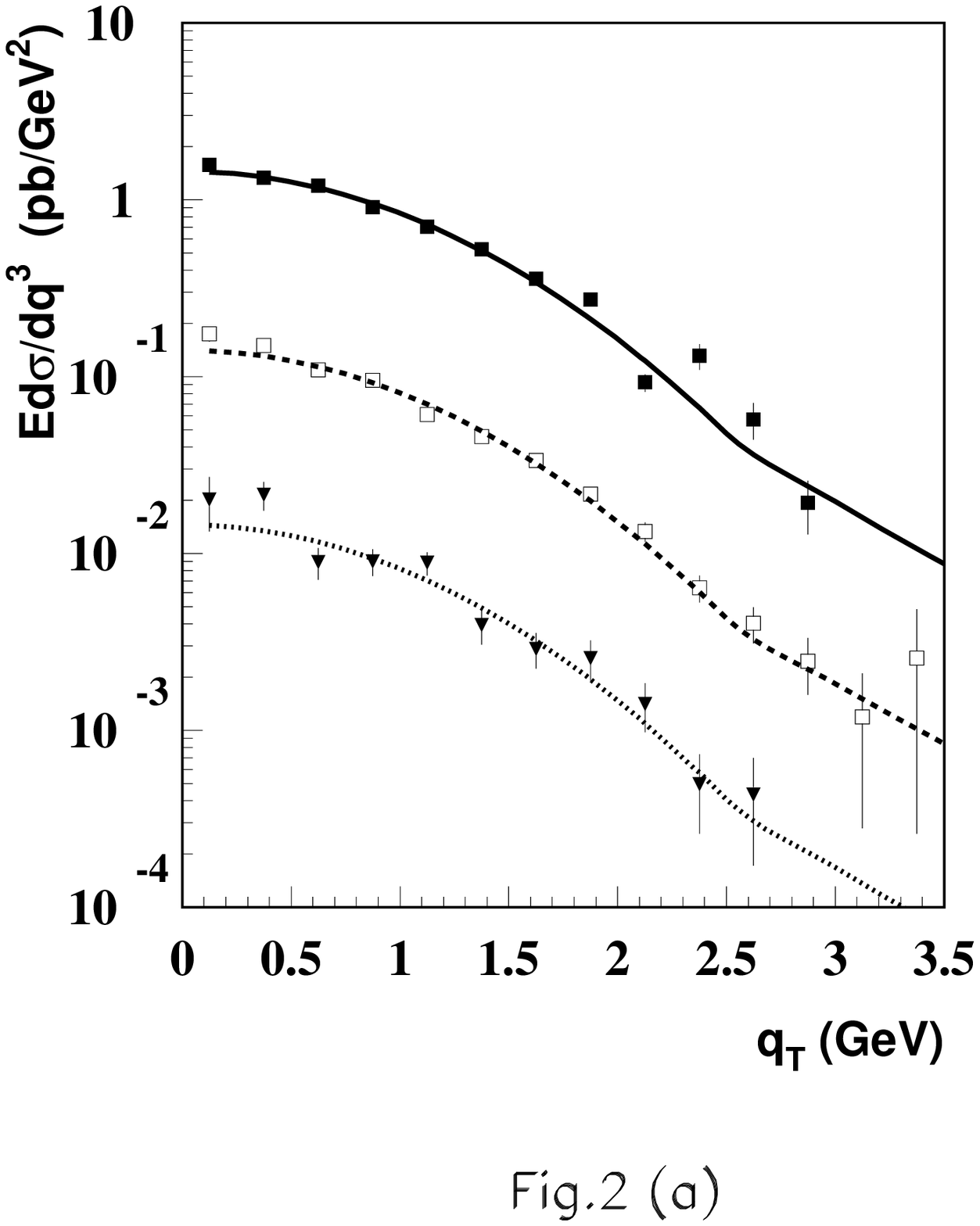,width=1.05in}
\end{minipage}
\hfill
\begin{minipage}[t]{2.0in}
\epsfig{figure=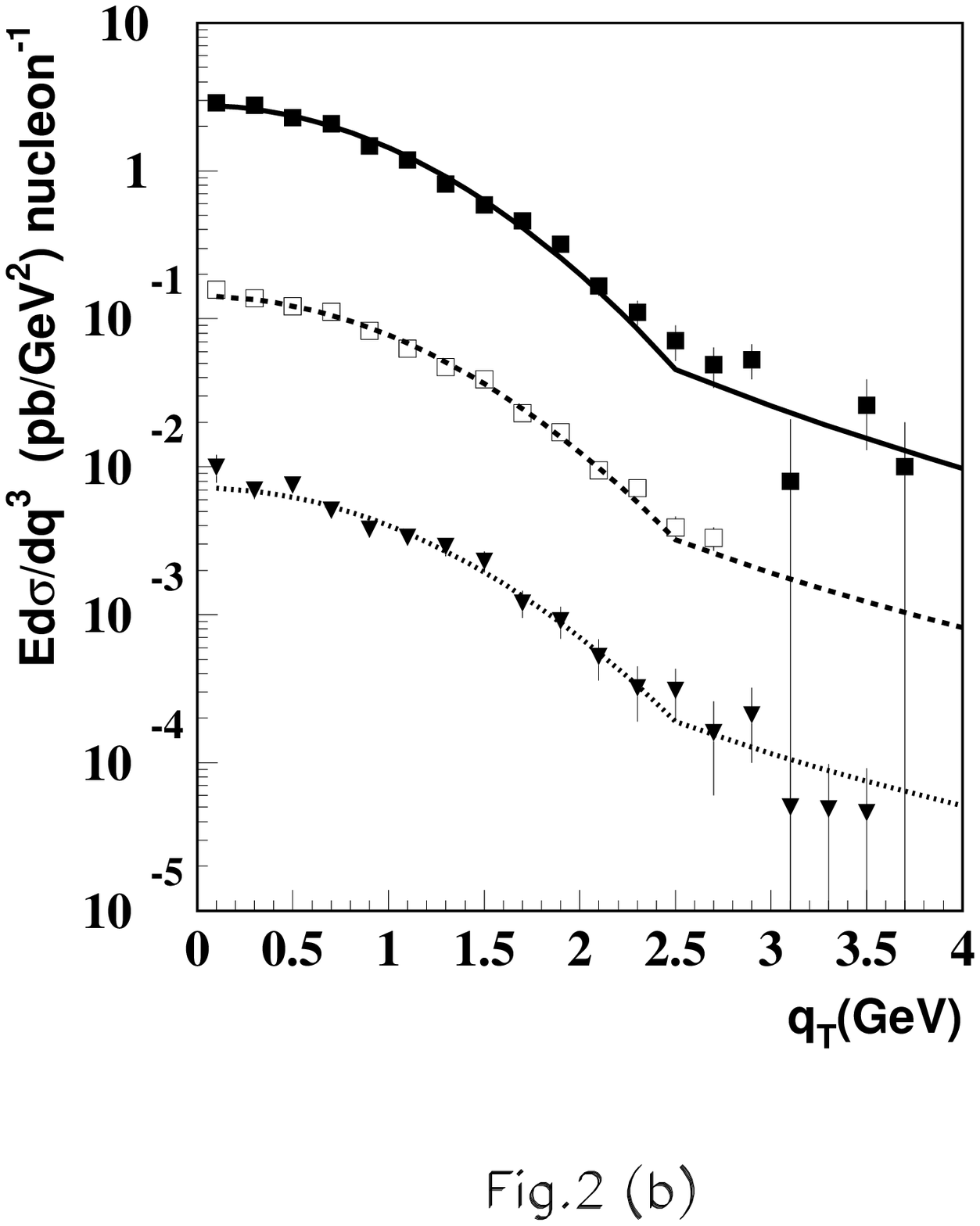,width=1.05in}
\end{minipage}
\caption{Data on the Drell-Yan transverse momentum spectrum: (a) E772
\protect\cite{DY-800} and (b) CFS \protect\cite{CFS}, in comparison
with a Gaussian-like fit in the small $q_T$ region plus a calculated
perturbative tail in the large $q_T$ region.}
\label{fig2}
\end{figure}

As argued in Ref.~\cite{CSS-resum}, the nonperturbative $F^{NP}(b)$
should have a Gaussian form in the $b$-space.  Since large $b$ values
correspond to a small $q_T$ region after the Fourier transform, 
the Gaussian form of $F^{NP}(b)$ is consistent with our $q_T$
spectrum in region (I).  Furthermore, the Drell-Yan $q_T$ spectrum 
in region (II) at fixed
target energies is mainly determined by the resummed $b$-space
distribution in large $b$ region, which is completely dominated by the
shape of $F^{NP}(b)$.  Consequently, the resummation performed in the
small $b$ region does not have any noticeable effect on the low $q_T$
spectrum at fixed target energies.  Experimentally, all
existing data for the Drell-Yan continuum between J/$\psi$ and
$\Upsilon$ peak at fixed target energies \cite{NA10,E772} can be well
represented by the Gaussian-like $q_T$ distribution for $q_T<q_T^L$.   
In Fig.~\ref{fig2}, we show that the Drell-Yan data at both
800~GeV \cite{DY-800} and 400~GeV \cite{CFS} can be
well represented by a Gaussian-like fit for $q_T$ up to 2.5~GeV, and
$Q=5.5$, 8.5, and 11.5~GeV, respectively.  The widths of
the Gaussian-like fits in Fig.~\ref{fig2} are consistent with the
parameters in the $F^{NP}(b)$ in the resummed Drell-Yan spectra
\cite{Davies,Yuan}.   

Actually, this finding should not be too surprising because for a
typical Drell-Yan pair of $Q\sim 5$~GeV and $q_T\sim 1$~GeV at fixed
target energies, the ratio $Q^2/q_T^2$ is about $25$, which is much
smaller than the $8\times 10^3$ for a $Z^0$ production at the
collider energies.  On the other hand, when $Q^2$ is much larger than
the typical $q_T^2$, the $q_T$ spectrum in this region 
has an enhanced peak at a finite value of $q_T$ and a slower fall off
in comparison with the Gaussian-like distribution, as sketched in
Fig.~\ref{fig1}.  The size of the enhancement and the size of the
region depend on the value of $Q$ and the range of $q_T$
\cite{Yuan}.  Furthermore, because of the slower fall of the
transverse momentum distribution with a full resummation of the large
logarithms at low $q_T$, the transition between the resummed
distribution and the large $q_T$ distribution calculated in the fixed
order perturbative QCD should be smoother than the unphysical kink, as
shown in Fig.~\ref{fig2}, at the point where the Gaussian-like
distributions are matched with the large $q_T$ perturbative tails.
Since we concentrate on Drell-Yan $q_T$ spectrum at
fixed target energies in this paper, we will effectively merge regions
(I) and (II) in the rest of our discussions.  

The large $q_T$ region, labeled by III in Fig.~\ref{fig1},  
corresponds to the region where $q_T> q_T^L$.  In this region, 
the transverse momentum spectrum calculated
in conventional fixed order perturbative QCD should be reliable
\cite{Berger,DY-LPT}, as shown by the solid line in Fig.~\ref{fig1}.  
This perturbative tail of
$q_T$-spectrum has a typical power-like behavior.  

In conclusion, the Drell-Yan $q_T$-spectrum at fixed target energies
can be represented by a Gaussian-like distribution in the small $q_T$
region and a perturbatively calculated tail in the large $q_T$ region,
\begin{eqnarray}
\frac{d\sigma}{dQ^2dq_T^2} &=&
  \frac{d\sigma^{(I)}}{dQ^2dq_T^2}\, \theta(q_T^L-q_T)
+ \frac{d\sigma^{(III)}}{dQ^2dq_T^2}\, \theta(q_T-q_T^L)
\nonumber \\
&=&
  \frac{d\sigma^{(I)}}{dQ^2dq_T^2}
+ \left(\frac{d\sigma^{(III)}}{dQ^2dq_T^2}
       -\frac{d\sigma^{(I)}}{dQ^2dq_T^2} \right) \theta(q_T-q_T^L)\, ,
\label{DYqt}
\end{eqnarray}
where $\frac{d\sigma^{(III)}}{dQ^2dq_T^2}$ is the perturbatively
calculated $q_T$-spectrum \cite{Berger,DY-LPT}, which is valid
for region (III), and $\frac{d\sigma^{(I)}}{dQ^2dq_T^2}$ 
is the Gaussian-like distribution defined in Eq.~(\ref{DYGau}), which 
is able to fit data in regions (I)+(II). 
Since $\frac{d\sigma^{(III)}}{dQ^2dq_T^2}$ is calculable, $q_T^L$
and parameters $N_{DY}$ and $\tau$ in Eq.~(\ref{DYGau}) are the only
unknowns for the Drell-Yan $q_T$-spectrum at fixed target energies.

Taking the moments of the Drell-Yan $q_T$ spectrum in
Eq.~(\ref{DYqt}), we can express the parameters $N_{DY}$ and $\tau$ in
Eq.~(\ref{DYGau}) in terms of physically measurable quantities.
For example, by integrating the $q_T$ spectrum in Eq.~(\ref{DYqt}), we
obtain the Drell-Yan total cross section as the zeroth moment of the $q_T$
spectrum. Using the data shown in Fig.~\ref{fig2}, we found that the
contribution from the second term in Eq.~(\ref{DYqt})
$\int_{q_T^L} \left(\frac{d\sigma^{(III)}}{dQ^2dq_T^2}
-\frac{d\sigma^{(I)}}{dQ^2dq_T^2}\right) dq_T^2$ is much less than one 
percent of the contribution from the first term 
$\int \left(\frac{d\sigma^{(I)}}{dQ^2dq_T^2} \right) dq_T^2$.
Therefore, up to the less than one percent uncertainty, we obtain the
normalization $N_{DY}$ in Eq.~(\ref{DYGau}) as 
\begin{equation}
N_{DY} \approx \frac{d\sigma}{dQ^2}\, .
\label{DYGau1}
\end{equation}

Define the Drell-Yan averaged transverse momentum square as
\begin{equation}
\langle q_T^2 \rangle \equiv
\frac{\int q_T^2 \left(d\sigma/dQ^2dq_T^2\right) dq_T^2}
     {d\sigma/dQ^2}\, .
\label{qt2}
\end{equation}
Substituting Eqs.~(\ref{DYqt}) and (\ref{DYGau1}) into our definition in
Eq.~(\ref{qt2}), we obtain
\begin{equation}
\langle q_T^2 \rangle = 2\tau^2 +  \Gamma(q_T^L,\tau^2)\ ,
\label{qt2-tau}
\end{equation}
with
\begin{equation}
\Gamma(q_T^L,\tau^2)\equiv 
\frac{1}{d\sigma/dQ^2} \int_{q_T^L}\, q_T^2 \left(
    \frac{d\sigma^{(III)}}{dQ^2dq_T^2} 
  - \frac{d\sigma^{(I)}}{dQ^2dq_T^2} \right) dq_T^2 \ . 
\label{Gamma-def} 
\end{equation}
In principle, $\Gamma(q_T^L,\tau^2)$  is calculable, and its $\tau^2$
dependence is from the $d\sigma^{(I)}/dQ^2dq_T^2$.   
Even though $\langle q_T^2\rangle$ enhances the contributions from the
perturbative tail, we find that the $\Gamma(q_T^L,\tau^2)$ in 
Eq.~(\ref{qt2-tau}) is much less than ten percent of the first term
for the data in Fig.~\ref{fig2} at fixed target energies. Therefore,  
by iteration, we derive the width of the Gaussian-like distribution
$d\sigma^{(I)}/dQ^2dq_T^2$ in Eq.~(\ref{DYGau}) as 
\begin{equation}
2\tau^2 \approx \langle q_T^2 \rangle 
- \Gamma(q_T^L, \tau^2=\langle q_T^2 \rangle /2)\, ,
\label{tau2}
\end{equation}
where the precise value of $q_T^L$ will be determined later.

Substituting Eqs.~(\ref{DYGau1}) and (\ref{qt2-tau}) 
into Eq.~(\ref{DYGau}),
we obtain the Drell-Yan $q_T$-spectrum at small $q_T$ 
in hadron-nucleon collisions as 
\begin{equation}
\frac{d\sigma^{hN}_{DY}}{dQ^2 dq_T^2} = 
\frac{d\sigma^{hN}_{DY}}{dQ^2} \,
\frac{1}{\langle q_T^2\rangle^{hN}_{DY}-\Gamma(q_T^L)^{hN}_{DY}} \,
{\rm e}^{-q_T^2/ (\langle q_T^2 \rangle^{hN}_{DY}
-\Gamma(q_T^L)^{hN}_{DY})}\ ,
\label{qtexpN}
\end{equation}   
where the superscript $hN$ represents the hadron-nucleon collisions, 
$\langle q_T^2 \rangle^{hN}_{DY}$ is the averaged transverse   
momentum square for the Drell-Yan production in hadron-nucleon 
collisions, and the short-handed notation $\Gamma(q_T^L)^{hN}_{DY}
\equiv 
\Gamma(q_T^L, \tau^2=\langle q_T^2 \rangle^{hN}_{DY}/2)^{hN}_{DY}$
with $\Gamma(q_T^L,\tau^2)$ defined in Eq.~(\ref{Gamma-def}).
We emphasize that all quantities, $d\sigma^{hN}/dQ^2$, $\langle q_T^2 
\rangle^{hN}_{DY}$, and $\Gamma(q_T^L)^{hN}_{DY}$, which completely
define the Drell-Yan $q_T$-spectrum in the small $q_T$ region, are either
independently measurable or perturbatively calculable.  
For the Drell-Yan data in Fig.~\ref{fig2} at fixed target energies, we
found that $\langle q_T^2 \rangle^{hN}_{DY}$ ranges from  
1.5~GeV$^2$ to 1.8~GeV$^2$, and $\Gamma(q_T^L)^{hN}_{DY}$ is less or
about ten percent depending on the value of $q_T^L$.

In order to derive an expression for the nuclear dependence
coefficient $\alpha(A,q_T)$, we need to know the Drell-Yan transverse
momentum spectrum in hadron-nucleus collisions.  In hadron-nucleus
collision, initial-state partons have much larger probability to 
have multiple scattering before the hard collision to produce the
Drell-Yan pair. Therefore, in order to obtain the transverse momentum  
spectrum in hadron-nucleus collisions,  we need to consider
contributions from  both the single scattering and the multiple
scattering.  For the single scattering contribution to the transverse
momentum spectrum in hadron-nucleus collisions, we can use the same
arguments given above for the spectrum in hadron-nucleon collisions.
Therefore, at fixed target energies, the {\it single} scattering
contribution to  $d\sigma^{hA}/dQ^2 dq_T^2$ in the small $q_T$ region
can be well represented by a Gaussian-like distribution. 

For the Drell-Yan production, parton initial-state interactions dominate 
the multiple scattering because the virtual photon does not interact 
strongly. The initial-state multiple scattering can provide extra
transverse momentum to a parton participating the hard collision, and 
consequently change the transverse momentum spectrum of the
produced Drell-Yan pair. Therefore, multiple scattering leads to a
broadening in the averaged transverse momentum square
$\langle q_T^2 \rangle ^{hA}$, which is defined in Eq.~(\ref{qt2})
with the superscript $hA$ representing the hadron-nucleus collisions.
Define  
\begin{equation}
\Delta \langle q_T^2\rangle \equiv 
\langle q_T^2 \rangle ^{hA}
- \langle q_T^2 \rangle^{hN} \, , 
\label{dydqt2}
\end{equation}
and $\Delta \langle q_T^2\rangle$ is often called  the
nuclear broadening of the averaged transverse momentum square. It was
demonstrated in Refs.~\cite{Guo2,Guo3} that the nuclear broadening 
$\Delta \langle q_T^2\rangle$ is calculable within QCD
perturbation theory.  In terms of the double scattering and the
generalized factorization theorem \cite{QS-fac}, the transverse
momentum broadening for the Drell-Yan pairs, $\Delta \langle
q_T^2\rangle_{DY}$, was calculated at the leading order in
$\alpha_s$, and was expressed in terms of the quark-gluon correlation
functions inside a nucleus \cite{Guo2}:
\begin{equation}
\Delta \langle q_T^2 \rangle_{DY} 
=\left(\frac{4\pi^2 \alpha_s}{3} \right)\cdot
\frac{\sum_{q} \, e_q^2\int dx' \, \phi_{\bar{q}/h}(x')\, 
T_{qg}^{SH}(Q^2/sx') /x'}
{\sum_{q}\, e_q^2 \int dx' \, \phi_{\bar{q}/h}(x')\, 
\phi_{q/A}(Q^2/sx') /x'} \, ,
\label{dyqt2b}
\end{equation}
where $\sum_q$ runs over all quark and antiquark flavors, $e_q$ is the 
fractional charge of quark and antiquark of flavor $q$, and $\alpha_s$
is the strong coupling constant.  In Eq.~(\ref{dyqt2b}), the
$s=(p+p')^2$ with $p'$ the four momentum of the incident hadron  
and $p$ the averaged momentum per nucleon for the nucleus, and 
$\phi_{\bar{q}/h}$ and $\phi_{q/A}$ are parton 
distributions of the hadron and the nucleus, respectively.  The 
$T_{qg}^{SH}(x)$ of flavor $q$ in Eq.~(\ref{dyqt2b}) is the
``soft-hard'' quark-gluon correlation function in a nucleus and is
defined by \cite{LQS1,Guo2}  
\begin{eqnarray}
T_{qg}^{SH}(x) &=&
 \int \frac{dy^{-}}{2\pi}\, e^{ixp^{+}y^{-}}
 \int \frac{dy_1^{-}dy_{2}^{-}}{2\pi} \,
      \theta(y^{-}-y_{1}^{-})\,\theta(-y_{2}^{-}) \nonumber \\
&\ & \times \,
     \frac{1}{2}\,
     \langle P_{A}|F_{\alpha}^{\ +}(y_{2}^{-})\bar{\psi}_{q}(0)
                  \gamma^{+}\psi_{q}(y^{-})F^{+\alpha}(y_{1}^{-})
     |P_{A} \rangle \ ,
\label{dyTq}
\end{eqnarray}
where we assumed that the nucleus is moving along the $+z$ direction,
and the superscripts ``+'' and ``$-$'' represent the normal $+$ and
$-$ components in a light-cone coordinate, respectively.  In 
Eq.~(\ref{dyTq}), the $y$, $y_1$ and $y_2$ represent the position
coordinates of the quark and gluon field operators.
Although these quark-gluon correlation functions are nonperturbative
and unknown, they are universal and as fundamental as the well-known
parton distributions \cite{LQS2}.  They appear not only in
Eq.~(\ref{dyqt2b}) for the nuclear broadening 
$\Delta \langle q_T^2 \rangle_{DY}$, but also in the nuclear dependence 
of the Drell-Yan
$q_T$-spectrum in hadron-nucleus collisions in the large $q_T$ region
\cite{Guo1,BMuller}.  In principle, by using the nuclear dependence of the 
Drell-Yan $q_T$-spectrum in the large
$q_T$ region, we can extract these quark-gluon correlation functions 
and predict the nuclear broadening 
$\Delta \langle q_T^2 \rangle_{DY}$, or vise versa.

On the other hand, by comparing the operator definitions of the
quark-gluon correlation functions $T^{SH}_{qg}(x)$, defined in
Eq.~(\ref{dyTq}), with the definitions of normalized quark
distributions,  
\begin{equation}
q^A(x)\equiv \frac{1}{A}\, \phi_{q/A}(x)\,
= \frac{1}{A}\, \int \frac{dy^-}{2\pi}\, {\rm e}^{ixp^+y^-}\,
\langle P_A|\bar{\psi}_q(0) \frac{\gamma^+}{2} \psi_q(y^-)
|P_A \rangle\, ,
\label{qA}
\end{equation}
one finds that the only difference between the $T^{SH}_{qg}(x)$ and
$q^A(x)$ is an operator factor,
$$
\int \frac{dy_1^-\, dy_2^-}{2\pi}\, \theta(y^- - y_1^-)\,
\theta(-y_2^-)\, F^{+}_{\ \alpha}(y_2^-)\,
                F^{+\alpha}(y_1^-)\, .
$$
If these two gluon field strength $F$'s are close together in a color
singlet nucleon in a normal color confined nuclear target, 
which limits the $\int dy_1^-$ (or $\int dy_2^-$) to a nucleon size, 
the extra integration $\int dy_2^-$ (or $\int dy_1^-$) can be
extended to the nuclear size to give an extra $A^{1/3}$ factor 
to the $T^{SH}_{qg}(x)$ over $q^A(x)$.  Therefore, one can estimate
the size of the quark-gluon correlation functions by modeling them as 
\cite{LQS1,LQS2}:
\begin{equation}
T_{qg}^{SH}(x)=\lambda^2\, A^{4/3} q^A(x) \ ,
\label{TiM-sh}
\end{equation}
where $\lambda$ is a universal parameter proportional to the averaged 
color field strength square in position space in a normal nuclear
target, and the $q^A(x)$ are the effective nuclear quark distributions
normalized by the atomic weight $A$.
With this model for $T_{qg}^{SH}$, the Drell-Yan transverse 
momentum broadening $\Delta \langle q_T^2\rangle_{DY}$ can be 
expressed as \cite{Guo2} 
\begin{equation}
\Delta \langle q_T^2 \rangle_{DY} = 
\left(\frac{4\pi^2\alpha_s}{3} \right) \
\lambda^2\ A^{1/3} \equiv b_{DY}\, A^{1/3} \ .
\label{dyqt2c}
\end{equation} 
The factor of $A^{1/3}$ in Eq.~(\ref{dyqt2c}) is an unique feature of
parton double scattering in a normal nuclear matter \cite{LQS2}.  
Data for the transverse momentum broadening is consistent with the 
features of double scattering \cite{E772,Guo2}.

Same as the single scattering case, double scattering contributions to 
the Drell-Yan $q_T$-spectrum in the small $q_T$ region is also
dominated by the intrinsic transverse momenta of scattering partons.
Similar to Eq.~(\ref{DYLO}), at the leading order in perturbation
theory, the {\it double} scattering contributions to the Drell-Yan 
$d\sigma^{hA}/dQ^2 dq_T^2$ 
can be described by \cite{Guo2}   
\begin{equation}
\frac{d\sigma^{hA}_D}{dQ^2 d^2q_T}
\propto T_{qg}(x,x_1,x_2,k_T) \, \delta^2(\vec{q}_T-\vec{k}_T)\ ,
\label{DO-spectrum}
\end{equation}
where the subscript $D$ indicates the double scattering, and the
proportionality coefficient depends on the structure of partonic double
scattering.  In Eq.~(\ref{DO-spectrum}), $\vec{q}_T$ is the transverse
momentum vector of the observed Drell-Yan pair, $\vec{k}_T$ represents
the intrinsic transverse momentum vector of the gluon which gives
additional scattering, and $T_{qg}(x,x_1,x_2,k_T)$
represents a general quark-gluon correlation function with $x$, $x_1$,
and $x_2$ being the light-cone momentum fractions carried by the quark
and gluon fields, and $k_T$ is the magnitude of $\vec{k}_T$
\cite{LQS2,Guo2}.  Integrations over parton momenta $dx\,dx_1\,dx_2$
reduce the general $T_{qg}(x,x_1,x_2,k_T)$ to either $T^{SH}_{qg}$,
defined in Eq.~(\ref{dyTq}), or $T^{DH}_{qg}$ which is defined
later in Eq.~(\ref{Tqg-DH}), depending on the structure of the 
partonic part of the double
scattering.  Similar to Eq.~(\ref{DYGau}), we can broaden the
$\delta$-function in Eq.~(\ref{DO-spectrum}) by its
Gaussian-representation to include the effect of intrinsic transverse
momentum of incoming quark or antiquark.  We can also take into
account the effect of gluon's intrinsic transverse momentum by using a 
Gaussian-like distribution for the $k_T$ dependence of the $T_{qg}$.
After integrating over the intrinsic $k_T$-dependence, we find that
just like the single scattering case, the double scattering
contributions to the Drell-Yan $q_T$-spectrum in the small $q_T$ region
can also be expressed in terms of a Gaussian-like distribution defined
in Eq.~(\ref{DYGau}) with a broadened width.
  
In order to understand the Drell-Yan $q_T$ spectrum in the region (II)
in hadron-nucleus collisions, in principle, we have to carry out the
resummation of the large logarithms at the presence of multiple
scattering.  Following the generalized factorization theorem
\cite{QS-fac}, the resummation of the large logarithms at the presence
of factorizable partonic multiple scattering can be carried out by
using renormalization group method.  In order to resum the large
logarithms due to gluon radiations, it is necessary to calculate
anomalous dimensions of the multiparton correlation functions for
deriving the kernel of the renormalization group equation.  However,
following the same arguments given above for the single scattering
case, the resummed Drell-Yan $q_T$-spectrum in region (II) at the
fixed target energies is mainly determined by the nonperturbative
input distribution $F^{NP}(b)$, which generalizes the perturbatively
resummed Drell-Yan $b$-space distribution to the nonperturbative large
$b$ regime in order to perform the Fourier transform back to the
$q_T$-space.  As argued above, the Drell-Yan $q_T$ spectrum at the
presence of multiple scattering should still have a Gaussian-like
distribution in the small $q_T$ region, and we expect that the
nonperturbative $F^{NP}(b)$ at the presence of multiple scattering
should also have a Gaussian-like distribution.  Therefore, at the
fixed target energies, we can merge the region (I) and (II) of the
Drell-Yan $q_T$-spectrum in hadron-nucleus collisions.  Since we limit 
this paper to test the consistency of QCD treatment of multiple
scattering at fixed target energies, which has not been done, we will
defer a complete analysis of QCD resummation at the presence of
multiple scattering to a future publication, which will be very
important for the nuclear dependence of Drell-Yan $q_T$ spectrum at
the collider energies.  

Similar to the case in hadron-nucleon scatterings, when 
$q_T > q_T^L$, the Drell-Yan $q_T$-spectrum should have a perturbatively 
calculable tail in region (III), and its nuclear dependences are 
perturbatively calculable and were calculated in Ref.~\cite{Guo1}.
Therefore, the Drell-Yan $q_T$-spectrum in hadron-nucleus collisions
should have the same form as that in Eq.~(\ref{DYqt}).
Using the calculated 
perturbative tail for the Drell-Yan spectrum in hadron-nucleus 
collisions, we find that similar to the hadron-nucleon case, the second 
term in Eq.~(\ref{DYqt}) contributes only about one percent of the total 
cross section $d\sigma^{hA}/dQ^2$.  Therefore, similar to
Eq.~(\ref{DYGau1}), we can identify the normalization factor of the
Gaussian-like distribution to be the total cross section 
$d\sigma^{hA}/dQ^2$.

Similar to Eq.~(\ref{qtexpN}), we derive
the Drell-Yan transverse momentum spectrum at small $q_T$ in
hadron-nucleus collisions as
\begin{equation}
\frac{d\sigma^{hA}}{dQ^2 dq_T^2} = \frac{d\sigma^{hA}}{dQ^2} \,
\frac{1}{\langle q_T^2\rangle^{hA}_{DY}-\Gamma(q_T^L)^{hA}_{DY}} 
{\rm e}^{-q_T^2/ (\langle q_T^2 \rangle^{hA}_{DY}
-\Gamma(q_T^L)^{hA}_{DY})} \ ,
\label{qtexpA}
\end{equation}   
where the superscript $hA$ represents the hadron-nucleus collisions, 
$\langle q_T^2 \rangle^{hA}_{DY} = \langle q_T^2
\rangle^{hN}_{DY} + \Delta\langle q_T^2 \rangle_{DY}$, and the
nuclear broadening $\Delta\langle q_T^2 \rangle_{DY}$ is given in
Eq.~(\ref{dyqt2b}) or Eq.~(\ref{dyqt2c}).  In Eq.~(\ref{qtexpA}),
$\Gamma(q_T^L)^{hA}_{DY}\equiv 
\Gamma(q_T^L, \tau^2=\langle q_T^2 \rangle^{hA}_{DY}/2)^{hA}_{DY}$
with $\Gamma(q_T^L,\tau^2)^{hA}_{DY}$ defined in Eq.~(\ref{Gamma-def})
with the calculable hadron-nucleon $q_T$-spectrum replaced by the 
corresponding hadron-nucleus $q_T$-spectrum.
Similar to the hadron-nucleon case, we find that the 
$\Gamma(q_T^L)^{hA}_{DY}$ in Eq.~(\ref{qtexpA}) is much smaller than
corresponding $\langle q_T^2 \rangle^{hA}_{DY}$.  
Since the Drell-Yan total cross section $d\sigma^{hA}/dQ^2$ 
has almost no nuclear dependence, Eq.~(\ref{qtexpA}) shows that the
nuclear broadening $\Delta\langle q_T^2\rangle_{DY}$ determines the
Drell-Yan $q_T$-spectrum at small $q_T$ in hadron-nucleus collisions. 

From Eqs.~(\ref{qtexpN}) and (\ref{qtexpA}), we can derive the 
nuclear dependence coefficient $\alpha(A,q_T)$ for the Drell-Yan 
process in small $q_T$ region.

\section{The Nuclear dependence coefficient for the Drell-Yan
production} 
\label{sec:3}

In this section, we derive the nuclear dependence coefficient 
$\alpha(A,q_T)$ for the Drell-Yan production in the small $q_T$ region. 
We also show that the derived $\alpha(A,q_T)$ in the small $q_T$ region
is naturally connected to the perturbatively calculated
$\alpha(A,q_T)$ in the large $q_T$ region. 

Substituting Eqs.~(\ref{qtexpN}) and (\ref{qtexpA}) into 
Eq.~(\ref{R-qt}), we derive the $\alpha_{DY}(A,q_T)$ for the
Drell-Yan production in the small $q_T$ region:
\begin{eqnarray}
\alpha_{DY}(A,q_T) &=& 1+\, 
\frac{1}{\ln(A)} \left[ \ln\left(R^A_{DY}(Q^2)\right) + 
\ln\left(\frac{1}{1+\chi_{DY}}\right)  \right. 
\nonumber \\ 
&\ & {\hskip 0.6in} \left. 
 + \frac{\chi_{DY}}{1+\chi_{DY}}\, 
\frac{q_T^2}{\langle q_T^2 \rangle^{hN}_{DY}-\Gamma(q_T^L)^{hN}_{DY}} 
\right]\, ,
\label{alpha-qt}
\end{eqnarray}
where $R^A_{DY}(Q^2)\equiv (1/A)(d\sigma^{hA}/dQ^2) / 
(d\sigma^{hN}/dQ^2)$.
The $\chi_{DY}$ in Eq.~(\ref{alpha-qt}) is defined by
\begin{equation}
\chi_{DY} \equiv 
\frac{\Delta \langle q_T^2 \rangle_{DY}
     -\Delta\Gamma(q_T^L)_{DY}} 
     {\langle q_T^2 \rangle^{hN}_{DY}-\Gamma(q_T^L)^{hN}_{DY}} 
\approx 
\frac{\Delta \langle q_T^2 \rangle_{DY}}
{\langle q_T^2 \rangle^{hN}_{DY}},
\label{chiDY}
\end{equation}
with $\Delta \langle q_T^2 \rangle_{DY}$ given in 
Eq.~(\ref{dyqt2c}), and 
$\Delta\Gamma(q_T^L)_{DY}\equiv \Gamma(q_T^L)^{hA}_{DY} -
\Gamma(q_T^L)^{hN}_{DY}$.  Because $\Gamma(q_T^L)_{DY}$ contributes 
less than ten percent to $\langle q_T^2 \rangle_{DY}$ in both
hadron-nucleon and hadron-nucleus collisions, 
$\Delta\Gamma(q_T^L)_{DY}$ is also much smaller than
$\Delta\langle q_T^2 \rangle_{DY}$.
Since the $\chi_{DY}$ and $\langle q_T^2\rangle^{hN}_{DY}$ 
have no dependence on $q_T$, the nuclear dependence
coefficient $\alpha_{DY}(A,q_T)$ in Eq.~(\ref{alpha-qt}) for the 
Drell-Yan production in the small $q_T$ region shows an universal
quadratic dependence on $q_T$:\ $\alpha_{DY}(A,q_T)=
a_1+b_1\,q_T^2$.  The parameters $a_1$ and $b_1$ 
are completely determined 
by either perturbatively calculable quantities (such as $\Delta\langle 
q_T^2\rangle$ and the $\Gamma(q_T^L)$'s) or independently
measurable quantities (such as $R^A_{DY}(Q^2)$ and $\langle q_T^2
\rangle^{hN}$). 

Using E772~\cite{E772} and NA10~\cite{NA10} data on $\Delta\langle
q_T^2 \rangle_{DY}$ and Eq.~(\ref{dyqt2c}), 
we obtain that $\Delta\langle q_T^2 \rangle_{DY} 
\sim 0.022A^{1/3}$~GeV$^2$, which is between 0.05 and 
0.13~GeV$^2$ for all practical nuclear targets.  On the other hand,
the value of $\langle q_T^2 \rangle^{hN}$ for the data in
Fig.~\ref{fig2} is ranging from 1.5 to 1.8~GeV$^2$.  Therefore, 
the $\chi_{DY}$ defined in Eq.~(\ref{chiDY}) should be a small number.

Substituting Eqs.~(\ref{dyqt2c}) and (\ref{chiDY}) into
Eq.~(\ref{alpha-qt}), and taking the small $\chi_{DY}$ limit, we obtain
\begin{eqnarray}
\alpha_{DY}(A,q_T) & \approx &
1+\, \frac{\chi_{DY}}{\ln(A)}\, \left[
-1 + \frac{q_T^2}{\langle q_T^2 \rangle^{hN}_{DY}-
\Gamma(q_T^L)^{hN}_{DY}} \right] 
+O\left(\frac{\chi_{DY}^2}{\ln(A)}\right)
\nonumber \\
& \approx &
1+\, \frac{b_{DY}}{\langle q_T^2 \rangle^{hN}_{DY}}\, \left[
-1 + \frac{q_T^2}{\langle q_T^2 \rangle^{hN}_{DY}} \right] \, ,
\label{alpha-qt0}
\end{eqnarray}
where $b_{DY}=(4\pi^2\alpha_s/3)\lambda^2$ is defined in
Eq.~(\ref{dyqt2c}). Here we used the experimental fact that
$R^A_{DY}(Q^2) \approx 1$. In deriving the second line in
Eq.~(\ref{alpha-qt0}), we used 
$A^{1/3}\sim \ln(A)$, which is a good approximation for most relevant
targets, and we neglected the small $\Gamma(q_T^L)^{hN}_{DY}$. 
 For a given beam energy, $\langle q_T^2 \rangle^{hN}_{DY}$ for
the Drell-Yan pairs has a very week dependence on $Q$ and no
dependence on $A$. Therefore, Eq.~(\ref{alpha-qt0}) shows that the
leading contribution to the $\alpha_{DY}(A,q_T)$ for the Drell-Yan
production is a function of $q_T$ and independent of $A$ in the small
$q_T$ region.    

From the size of the correction term in Eq.~(\ref{alpha-qt0}), it is
clear that the $\alpha_{DY}(A,q_T)$ for the Drell-Yan production is 
insensitive to the atomic weight $A$ of targets.  Furthermore, from
Eq.~(\ref{alpha-qt}), we obtain 
\begin{eqnarray}
\alpha_{DY}(q_T=0)  &\approx & 1 - \frac{\ln(1+\chi_{DY})}{\ln(A)}
\nonumber \\
&\approx &
1- \frac{\chi_{DY}}{\ln(A)} \approx 
1 -  \left(\frac{4\pi^2\alpha_s}{3} \right) \, 
\frac{\lambda^2}{\langle q_T^2 \rangle^{hN}_{DY}} \, .
\label{alpha0}
\end{eqnarray}
From Eqs.~(\ref{chiDY}) and (\ref{alpha0}), it is clear that 
$\alpha_{DY}(q_T=0)$ is determined by the nuclear broadening 
$\Delta \langle q_T^2 \rangle_{DY}$.  If there is nuclear broadening
in averaged transverse momentum square (i.e., $\chi_{DY}>0$), the
transverse momentum spectrum is suppressed at $q_T\sim 0$.  For a
normal nuclear target, the amount of the suppression at $q_T\sim 0$
is insensitive to the atomic number $A$ of targets. 

Eq.~(\ref{alpha-qt}) shows that as $q_T$ increases,
$\alpha_{DY}(A,q_T)$ increases as well, and it changes from less than
one to a value larger than one.  Let $q_T^c$ correspond to the
critical value of $q_T$ at which $\alpha_{DY}(A,q_T^c)=1$, and we 
have 
\begin{equation}
(q_T^c)^2 \approx  \langle q_T^2 \rangle^{hN}_{DY}\,
          \left[\frac{1+\chi_{DY}}{\chi_{DY}}\right]\,
            \ln(1+\chi_{DY})
\approx \langle q_T^2 \rangle^{hN}_{DY}\,
        \left[1+O(\chi_{DY})\right]\, .
\label{qtc}
\end{equation}
Like $\alpha_{DY}(A,q_T=0)$, the $q_T^c$ in Eq.~(\ref{qtc}) has a very 
weak dependence on the atomic number $A$ for a normal nuclear target.

In the rest of this section, we show that the $\alpha_{DY}(A,q_T)$
in Eq.~(\ref{alpha-qt}) (or approximately in Eq.~(\ref{alpha-qt0}))
for the small $q_T$ region is naturally connected to the perturbatively
calculated $\alpha_{DY}(A,q_T)$ in the large $q_T$ region \cite{Guo1}. 

\begin{figure}
\epsfig{figure=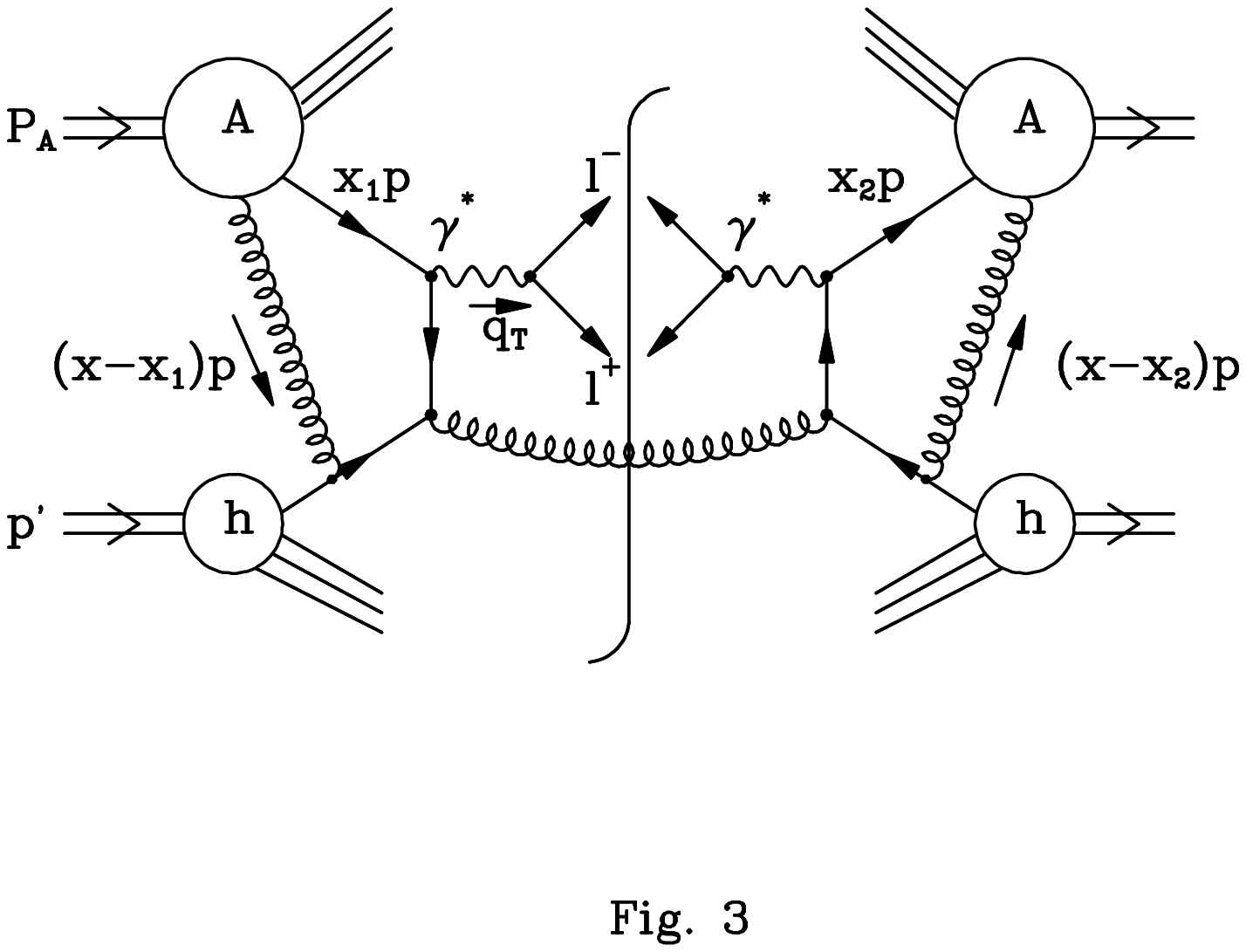,width=1.1in}
\caption{Sample Feynman diagram for the double scattering
contributions to the nuclear dependence of the Drell-Yan $q_T$-spectrum   
at large $q_T$.} 
\label{fig3}
\end{figure}

The nuclear dependence of the Drell-Yan transverse
momentum spectrum in the large $q_T$ region can be calculated in QCD
perturbation theory \cite{Guo1,BMuller}.  The ratio of the Drell-Yan
$q_T$-spectrum $R(A,q_T)=A^{\alpha(A,q_T)-1}$ defined in
Eq.~(\ref{R-qt}) was derived in  Ref.~\cite{Guo1} for the large $q_T$
region.  It was found \cite{Guo1} that at the leading order in $\alpha_s$,
the $R(A,q_T)$ depends on two types of parton correlation functions
due to different structure of partonic double scattering. 
For example, consider a double scattering Feynman diagram in 
Fig.~\ref{fig3}, which
contributes to the nuclear dependence of the Drell-Yan $q_T$-spectrum in
the large $q_T$ region.  In Fig.~\ref{fig3}, $x$, $x_1$, and $x_2$ are
independent parton momentum fractions for the double scattering.  
For the contour integration of $dx_i$'s, taking the residues of 
poles of the partonic double scattering diagram can eliminate the 
oscillation factor e$^{-x_ip^+ y_i^-}$ for corresponding integrations
$dy_i^-$'s in position space \cite{LQS1,Guo1}.  The lack of the
oscillation factor for integration of position variable $y_i^-$'s
leads to an $A^{1/3}$-type (or nuclear size) dependence.  The partonic
double scattering diagram in Fig.~\ref{fig3} has two types of poles:
one type that  corresponds  
to $x-x_1\rightarrow 0$ and $x-x_2\rightarrow 0$, and the other type that 
corresponds to $x_1=x_2\equiv x_a$ and $x-x_1=x-x_2\equiv x_b$ with
both $x_a$ and $x_b$ finite.  The first type poles lead to the
so-called ``soft-hard'' quark-gluon correlation functions defined in
Eq.~(\ref{dyTq}), which have both $dy_1^-$ and $dy_2^-$ integrations
free of the oscillation factor.  The second type poles lead to the
so-called ``double-hard'' quark-gluon correlation functions
\cite{LQS1,Guo1} 
\begin{eqnarray}
T_{qg}^{DH}(x_a,x_b) &=& \frac{1}{x_bp^+}
 \int \frac{dy_1^{-}}{2\pi}
 \int \frac{dy^{-}}{4\pi} \int p^+dy_{2}^{-} 
      \theta(y^{-}-y_1^{-})\theta(-y_{2}^{-}) 
        e^{ix_ap^{+}y^{-}} e^{ix_bp^+(y_1^--y_2^-)}\nonumber \\
&\ & {\hskip 0.3in} \times 
     \langle P_{A}|F_{\sigma}^{\ +}(y_{2}^{-})\bar{\psi}_{q}(0)
                  \gamma^{+}\psi_{q}(y^{-})F^{+\sigma}(y_1^{-})
     |P_{A}\rangle\, ,
\label{Tqg-DH} 
\end{eqnarray}
where all the variables and the superscripts carry the same meaning as
those in Eq.~(\ref{dyTq}).
Unlike the soft-hard correlation functions, only one $dy_2^-$ (or
$dy_1^-$) integration in the double-hard correlation functions defined
in Eq.~(\ref{Tqg-DH}) is free of the oscillation factor.  But, this
$dy_2^-$ (or $dy_1^-$) integration is just enough to produce an
$A^{1/3}$-type enhancement to the Drell-Yan $q_T$-spectrum 
\cite{LQS1,Guo1}. 
In addition to the double-hard quark-gluon correlation functions,
the nuclear dependence at large $q_T$  also receives contributions
from quark-quark and gluon-gluon correlation functions \cite{Guo1}.
By comparing the operator definitions of the normal quark and gluon
distributions with the operator definitions for the double-hard
correlation functions, similar to Eq.~(\ref{TiM-sh}), 
the double-hard quark-gluon correlation functions can be parameterized
as \cite{MQ}
\begin{equation}
T_{qg}^{DH}(x_a,x_b) =(2\pi C)\, A^{4/3}\, q^A(x_a)\, G^A(x_b) \ ,
\label{TiM-dh}
\end{equation}
with $q^A(x_a)$ and $G^A(x_b)$ the effective nuclear parton
distributions normalized by the atomic number $A$. 
The  constant parameter $C$ represents the
size of quantum interference between different nucleon states and
possible overlap of two nucleon parton distributions in a nucleus.
Currently, there is no direct observable yet to extract
information on $T^{DH}$ \cite{BMuller} or to determine the parameter 
$C$ in Eq.~(\ref{TiM-dh}). 

When $x_1$ (or $x_2$) goes to zero, the corresponding parton fields 
reach the saturation region, and the $T^{DH}$ is reduced to $T^{SH}$.
Therefore, if we assume the full quantum interference (in
saturation region), the proportionality constant $C$
in Eq.~(\ref{TiM-dh}) can be related to the $\lambda^2$ in
Eq.~(\ref{TiM-sh}) by using the following identity: 
\begin{equation}
\lim_{x_b \rightarrow 0} \, x_b T_{qg}^{DH}(x_a,x_b)
=T_{qg}^{SH}(x_a) \ ,
\label{Tsh-dh}
\end{equation}
From Eq.~(\ref{Tsh-dh}) and the definitions of $T^{DH}_{qg}$ and
$T^{SH}_{qg}$, we obtain 
\begin{equation}
C \approx \frac{\lambda^2}{2\pi xG(x)|_{x\approx 0}}\, , 
\label{c-lambda}
\end{equation}
where $xG(x)|_{x\approx 0}$ is the gluon momentum density at
$x\rightarrow 0$ limit, which is believed to be order of one
at low energy \cite{xG,Mueller}.  
The precise value of
$xG(x)|_{x\approx 0}$ depends on the physics of parton saturation
\cite{Mueller}, and it has not been well measured experimentally.
In this paper, we choose $xG(x)|_{x\approx 0}\approx 3$ based on the 
simple form $ G(x)\approx 3\,x^{-1}\,(1-x)^5$, which is a result of
a simple power counting for gluons at large $x$, Regge behavior for
gluons at small $x$, and an assumption that gluons carry 50 percent
momentum of the beam hadron. 
Different choice for 
$xG(x)|_{x\approx 0}$ results into a different overall normalization
factor $C$.
The $\lambda^2$ in Eq.~(\ref{c-lambda}) is the same
as the one in Eq.~(\ref{TiM-sh}) for the soft-hard quark-gluon
correlation functions.  Other correlation functions can also be 
determined in a similar way.    
Therefore, the $\lambda^2$ becomes the only parameter to fix the
normalization of both the soft-hard $T^{SH}$ and the double-hard $T^{DH}$
correlation functions.  If we extract the value of $\lambda^2$ from
the data on the transverse momentum broadening  
$\Delta\langle q_T^2 \rangle_{DY}$ by using Eq.~(\ref{dyqt2c}),
we can fix our predictions of the nuclear dependence for the Drell-Yan
$q_T$-spectrum in the large $q_T$ region.

On the other hand, by assuming that the quantum interference between 
parton field operators from different nucleon states are strongly
suppressed, one derives 
\begin{equation}
C=0.35/(8\pi r_0^2) \ {\rm GeV}^2 
\label{C-1}
\end{equation}
with $r_0 \approx 1.1 -1.25$~fm, which is just a geometric 
factor for finding two nucleons at 
the same impact parameter \cite{Guo1,MQ}.

Because of a combination of a small value of the measured $\lambda^2$
from Drell-Yan data \cite{NA10,E772} and a choice of
$[xG(x)]_{x\approx 0} \approx 3$, the numerical values of $C$
determined by Eqs.~(\ref{C-1}) and (\ref{c-lambda}) approximately
differ by a factor of 20.  We believe that the numerical value for the
$C$ given in Eq.~(\ref{C-1}) without any quantum interference between
parton field operators should represent a possible maximum for the
$C$, while the value given in Eq.~(\ref{c-lambda}) with full quantum
interference (in saturation region) represents a possible minimum for
the $C$. The parameter $C$ reflects the size of quantum interference
involved in the multiparton correlation functions, and its precise
value should be between those obtained by Eqs.~(\ref{C-1}) and
(\ref{c-lambda}). More discussions on the consequences of a different
value of the $C$ are given in Sec.~\ref{sec:5}.

\begin{figure}
\begin{minipage}[t]{2.0in}
\epsfig{figure=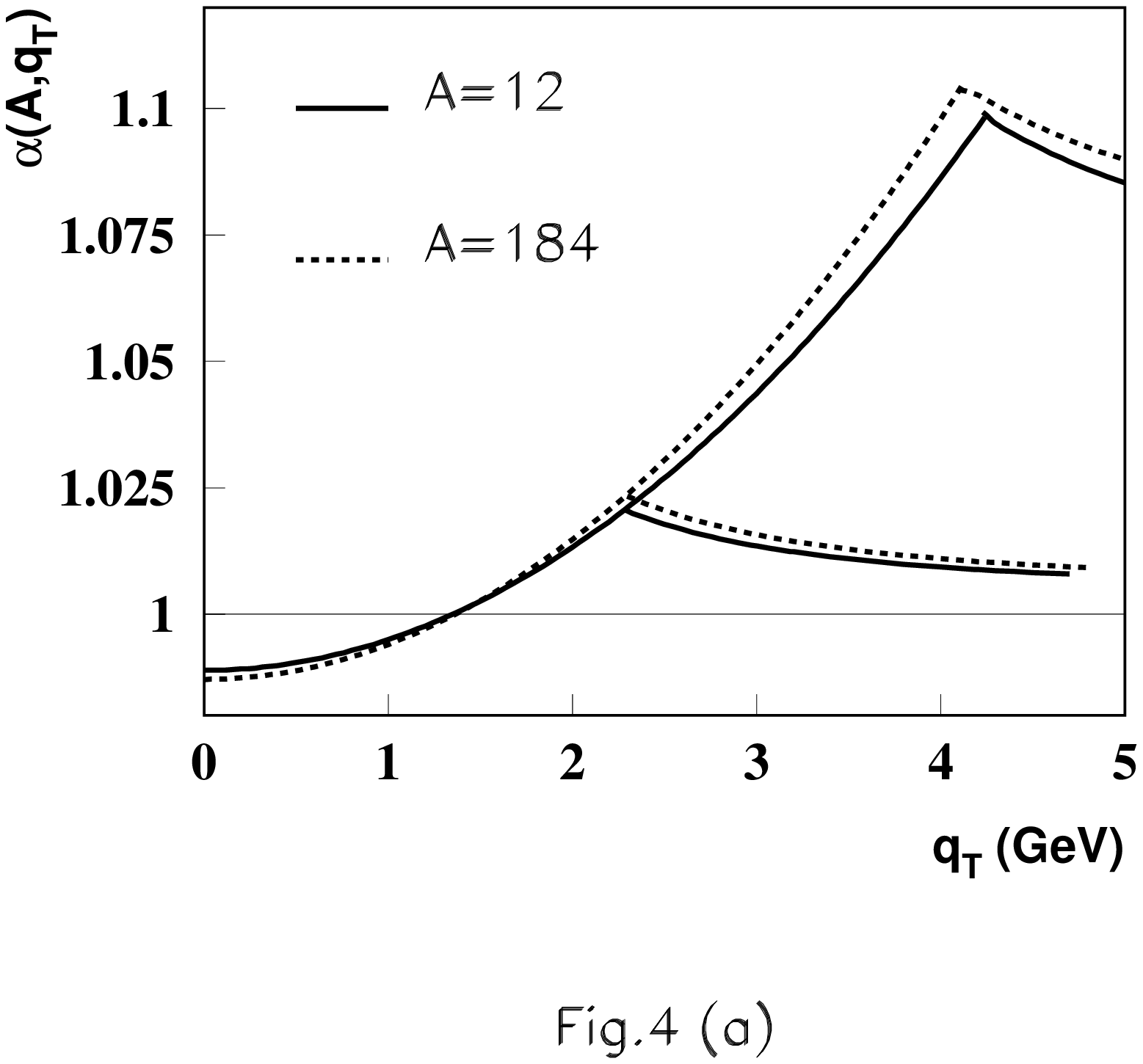,width=1.05in}
\end{minipage}
\hfill
\begin{minipage}[t]{2.0in}
\epsfig{figure=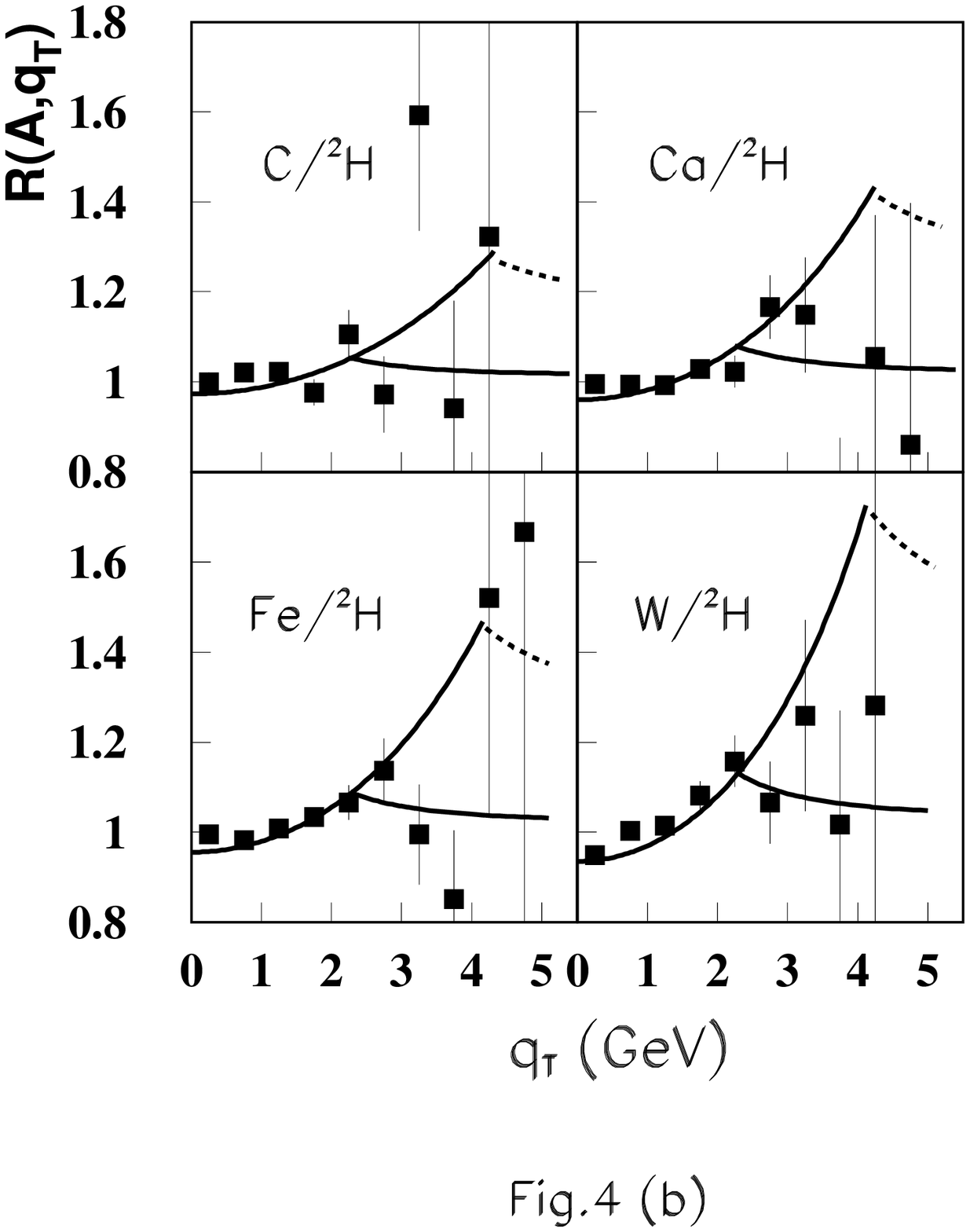,width=1.05in}
\end{minipage}
\caption{(a) $\alpha(A,q_T)$ for the Drell-Yan production as a
function of $q_T$.  Low $q_T$ region is given by 
Eq.~(\protect\ref{alpha-qt}), and high $q_T$ region is given by 
Ref.~\protect\cite{Guo1} with the maximum and minimum values of the 
$C$ discussed in the
text. (b) $R(A,q_T)$ for the Drell-Yan production
as a function of $q_T$.  Data from E772 \protect\cite{E772,E772-moss},
and the theory curves are given by $R(A,q_T)=A^{\alpha(A,q_T)-1}$ with 
$\alpha(A,q_T)$ given in (a).}
\label{fig4}
\end{figure}

In Fig.~\ref{fig4}, we plot the $\alpha_{DY}(A,q_T)$ 
as a function of $q_T$ for various nuclear targets.  For the
small $q_T$ region, we used our $\alpha_{DY}(A,q_T)$ in 
Eq.~(\ref{alpha-qt}). For the large $q_T$ region, we used the
perturbative calculations of $R(A,q_T)\equiv A^{\alpha(A,q_T)-1}$ in
Ref.~\cite{Guo1}.
In plotting Fig.~\ref{fig4}a, we used $\langle q_T^2
\rangle^{hN}_{DY}=1.8$~GeV$^2$, and $\Delta\langle q_T^2 
\rangle_{DY} = 0.022 A^{1/3}$~GeV$^2$.  
We choose the value of $\lambda^2$ to be consistent with 
$b_{DY}=0.022$~GeV$^2$.  The solid and dashed 
lines correspond to the nuclear target C ($A$=12) and 
W ($A$=184), respectively.  
For the large $q_T$ region, we plotted the perturbatively calculated
$\alpha_{DY}(q_T)$ \cite{Guo1}  
with the maximum and minimum values of the $C$
discussed above. 
It is clear from Fig.~\ref{fig4}a that our nuclear dependence
coefficient for the Drell-Yan $q_T$-spectrum in the small $q_T$ region 
is naturally connected to the perturbatively calculated
$\alpha_{DY}(A,q_T)$ in the large $q_T$ region.  Without any
adjustment of parameters, we choose the $q_T^L$ to be the transverse
momentum at the cross point between the low $q_T$ and high $q_T$
spectrum.  Without any extra free fitting
parameters, our predictions shown in Fig.~\ref{fig4} are consistent
with data in small $q_T$ region, and due to the large error in data
at high $q_T$, current Drell-Yan data at the fixed target energies are
consistent with almost any value of the $C$ between the maximum 
and the minimum discussed above.  Different value of $C$ resulted in
different value of $q_T^L$ at the crossing point. 
From Fig.~\ref{fig4}a, the value of $q_T^L$ is  between 
$2.3-4.0$~GeV, which is very reasonable.  

As expected from Eq.~(\ref{alpha-qt0}), the
$\alpha_{DY}(A,q_T)$ in Fig.~\ref{fig4}a shows very little
dependence on the atomic weight $A$.  Therefore, the
$\alpha_{DY}(A,q_T)$ for the Drell-Yan production in the small $q_T$
region has scaling on $A$ and shows an universal quadratic
dependence on $q_T$.  The scaling violation due to the potential
$A$-dependence is extremely small for a normal color confined nuclear
target.  Since $\langle q_T^2 \rangle^{hN}_{DY}$ can depend
on the beam energy and $Q$ as well as the scaled longitudinal 
momentum ($x_F$), our
universal function $\alpha_{DY}(A,q_T)$ in Eq.~(\ref{alpha-qt}) can
depend on these observables accordingly \cite{Guo3}. 

In Fig.~\ref{fig4}b, we compare our $\alpha_{DY}(A,q_T)$ shown in
Fig.~\ref{fig4}a with data from Fermilab experiment E772
\cite{E772,E772-moss}.  Using the $\alpha_{DY}(A,q_T)$ in 
Fig.~\ref{fig4}a, we plot the ratio of cross sections
$R(A,q_T)=A^{\alpha(A,q_T)-1}$ as a function of $q_T$.
All parameters are the same as those used in plotting
Fig.~\ref{fig4}a.  Also plotted in Fig.~\ref{fig4}b are data from E772
on four different nuclear targets: C, Ca, Fe, and W
\cite{E772,E772-moss}. It is very clear that 
without any extra fitting parameter, our predictions for the nuclear
dependence coefficient $\alpha_{DY}(A,q_T)$ provide a good description
of the data on $R(A,q_T)$ for all values of $q_T$.
The $A$-dependence in $R(A,q_T)$ in Fig.~\ref{fig4}b is clearly
consistent with the $A$-independence of $\alpha_{DY}(A,q_T)$ shown in
Fig.~\ref{fig4}a.  Better data can provide even more critical tests of
our predictions for $\alpha_{DY}(A,q_T)$.  

Fig.~\ref{fig4} demonstrates a clear consistency of the generalized
QCD factorization theorem \cite{QS-fac}, which was used to calculate
the nuclear broadening $\Delta \langle q_T^2 \rangle_{DY}$ and
the nuclear dependence in the large $q_T$ region.  Even though the
nuclear broadening $\Delta \langle q_T^2 \rangle_{DY}$ is 
dominated by information at low $q_T$ while the perturbative tail (or
the $\alpha_{DY}(A,q_T)$ in the large $q_T$ region) is controlled by 
dynamics at large $q_T$, our derived $\alpha_{DY}(A,q_T)$ for the small 
$q_T$ region is naturally linked to that for the large $q_T$ region 
without free fitting parameters.  More discussions on the
uncertainties, such as the resummation and the choice of the parameter
$C$, will be given in Sec.~\ref{sec:5}.

\section{The Nuclear dependence coefficient for 
         J/$\psi$ production}
\label{sec:4}

A major difference between the Drell-Yan production and J/$\psi$
production is the final-state interaction.  Because of the virtual
photon, the Drell-Yan production has only initial-state interactions while
J/$\psi$ production has both initial-state and final-state
interactions.  Therefore, one may conclude that the nuclear 
dependence of the transverse momentum spectrum for J/$\psi$ and
the Drell-Yan production could be different due to 
the strong final-state interactions. However, we argue in this section 
that in the small $q_T$ region at fixed target energies, J/$\psi$ 
transverse momentum spectrum can also be well represented by a
Gaussian-like distribution, and the nuclear dependence coefficient 
$\alpha(A,q_T)$ for J/$\psi$ production is very similar to that 
for the Drell-Yan production.

\begin{figure}
\epsfig{figure=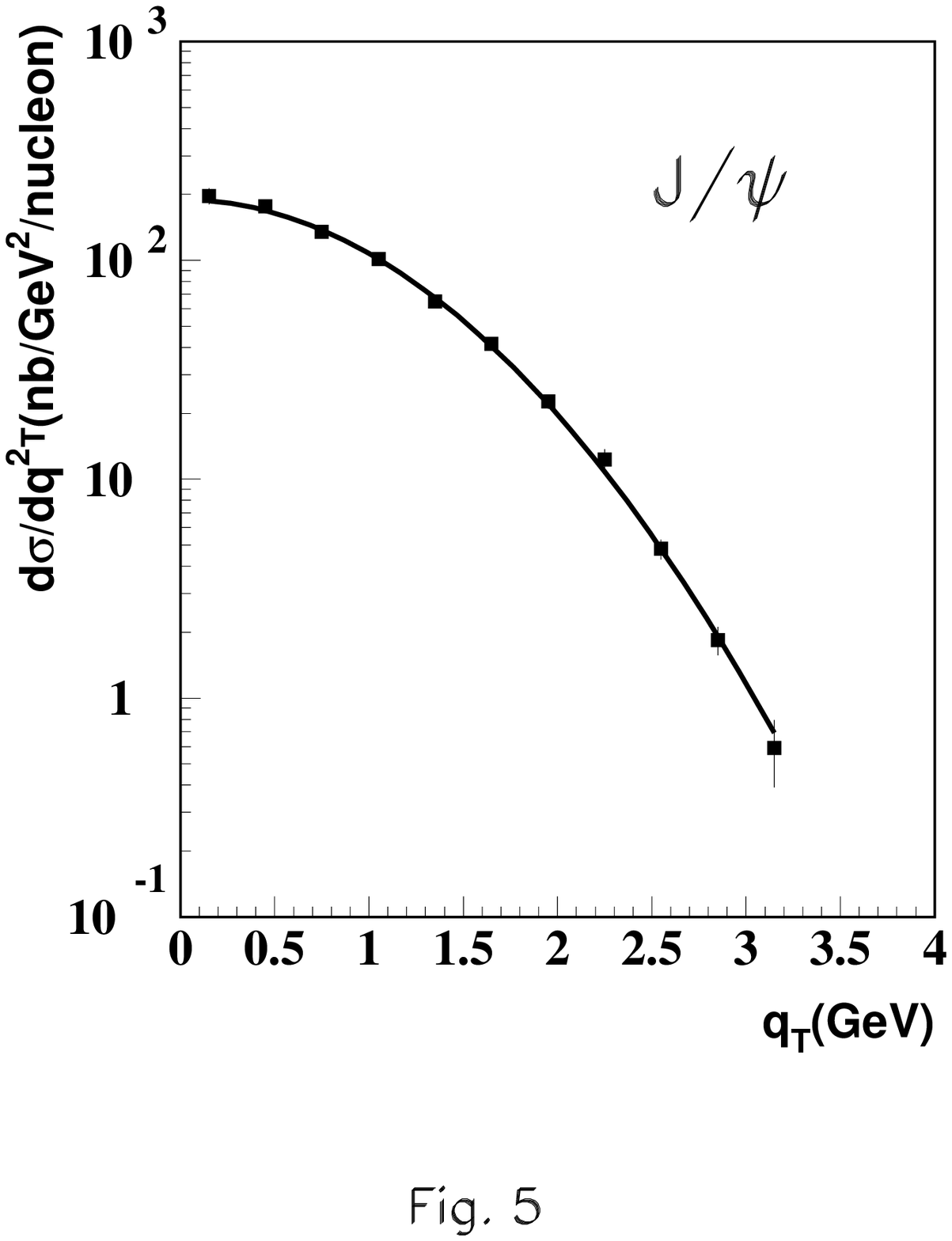,width=1.1in}
\caption{Fermilab E771 data on J/$\psi$ transverse momentum spectrum
\protect\cite{E771} in comparison with a fit of Gaussian-like
distribution.} 
\label{fig5}
\end{figure}

Kinematically, hadronic J/$\psi$ production is very much like
the Drell-Yan production with $Q\sim M_{{\rm J/}\psi}$.  Therefore, 
the J/$\psi$ transverse momentum spectrum also has three
similar characteristic regions shown in Fig.~\ref{fig1}.  However, 
because J/$\psi$ mass $M_{{\rm J/}\psi}$ is smaller than any typical
$Q$ measured from the Drell-Yan continuum, the logarithm $(\alpha_s
\ln^2(M_{{\rm J/}\psi}/q_T^2))^n$ for J/$\psi$ production should be
less important than any Drell-Yan production.  We then expect
that at fixed target energies, a Gaussian-like distribution can fit
J/$\psi$ transverse momentum spectrum at small $q_T$ even better.  In
Fig.~\ref{fig5}, we plot data on J/$\psi$ transverse momentum spectrum 
from Fermilab experiment E771 \cite{E771}.  The solid line in
Fig.~\ref{fig5} is a fit using a Gaussian-like distribution
\begin{equation}
\frac{d\sigma^{(G)}_{{\rm J/}\psi}}{dq_T^2}
=  N_{{\rm J/}\psi}\, \frac{1}{2\tau^2}\,
{\rm e}^{-q_T^2/2\tau^2} \, ,
\label{JPsi-fit}
\end{equation}
where superscript ``(G)'' stands for a Gaussian-like distribution, 
$N_{{\rm J/}\psi}$ and
$\tau$ are the normalization and the width of the Gaussian-like fit,
respectively.  As expected, the $\chi^2$ for fitting the J/$\psi$
spectrum in Fig.~\ref{fig5} is about 0.9 per degree of freedom, which
is much smaller than the $\chi^2$ for fitting Drell-Yan data in 
Fig.~\ref{fig2}, where the $\chi^2$ per degree of freedom is ranging
from 1.0 to 2.0.

Similar to Eqs.~(\ref{DYLO}) and (\ref{DO-spectrum}), at the
lowest order in perturbation theory, both the initial-state and
final-state double scattering contribution to the J/$\psi$ transverse
momentum spectrum in hadron-nucleus collisions is proportional a
$\delta$-function: $\delta^2(\vec{q}_T-\vec{k}_T)$, where $q_T$ is the
total transverse momentum of any produced pre-J/$\psi$ partonic state
(e.g., a state of charm and anticharm quark pair), and $k_T$
represents the intrinsic momentum of the gluon which gives additional
scattering.  Following the same arguments that follows 
Eq.~(\ref{DO-spectrum}) in Sec.~\ref{sec:2}, we conclude that in the small
$q_T$ region, J/$\psi$ transverse momentum spectrum in hadron-nucleus 
collisions can be well represented by a Gaussian-like
distribution shown in Eq.~(\ref{JPsi-fit}), and the normalization
$N_{{\rm J/}\psi}$ and the width $\tau$ are given by
\begin{eqnarray} 
N_{{\rm J/}\psi} &\approx & \sigma^{hA}_{{\rm J/}\psi} \, ,
\nonumber \\
2\tau^2 &\approx & \langle q_T^2\rangle^{hA}_{{\rm J/}\psi}-
\Gamma(q_T^L)^{hA}_{{\rm J/}\psi}\, ,
\label{JPsi-Ntau}
\end{eqnarray}
where $\sigma^{hA}_{{\rm J/}\psi}$ is the J/$\psi$ total cross
section, and $\langle q_T^2\rangle^{hA}_{{\rm J/}\psi}$ is the
averaged transverse momentum square, defined in the same way as that
for the Drell-Yan $\langle q_T^2\rangle^{hA}_{DY}$ in Eq.~(\ref{qt2}).
In Eq.~(\ref{JPsi-Ntau}), $\Gamma(q_T^L)^{hN}_{{\rm J/}\psi}$ is
perturbatively calculable and defined as
\begin{equation}
\Gamma(q_T^L)^{hA}_{{\rm J/}\psi} \equiv 
\frac{1}{\sigma^{hA}_{{\rm J/}\psi}}\, 
\int_{q_T^L}\, q_T^2 \left(
    \frac{d\sigma^{(P)}_{{\rm J/}\psi}}{dq_T^2} 
  - \frac{d\sigma^{(G)}_{{\rm J/}\psi}}{dq_T^2} \right) dq_T^2 \, ,
\label{Gamma-jpsi} 
\end{equation}
where $d\sigma^{(P)}_{{\rm J/}\psi}/dq_T^2$ is a
perturbatively calculable $q_T$-spectrum at large $q_T$, and 
$d\sigma^{(G)}_{{\rm J/}\psi}/dq_T^2$ is defined in
Eq.~(\ref{JPsi-fit}) with 
$2\tau^2=\langle q_T^2\rangle^{hA}_{{\rm J/}\psi}$.
The corresponding transverse momentum spectrum in the small $q_T$ region
is given by 
\begin{equation}
\frac{d\sigma_{{\rm J/}\psi}^{hA}}{dq_T^2}
= \sigma_{{\rm J/}\psi}^{hA} \, 
  \frac{1}{\langle q_T^2\rangle^{hA}_{{\rm J/}\psi}
-\Gamma(q_T^L)^{hA}_{{\rm J/}\psi}}\,
{\rm e}^{-q_T^2/(\langle q_T^2 \rangle^{hA}_{{\rm J/}\psi}
-\Gamma(q_T^L)^{hA}_{{\rm J/}\psi})} \, .
\label{JPsiexpA}
\end{equation}
The $\Gamma(q_T^L)^{hA}_{{\rm J/}\psi}$ is the small contribution 
to $\langle q_T^2\rangle^{hA}_{{\rm J/}\psi}$ due to the difference
between the perturbative tail and an extended Gaussian-like
distribution, and is defined in the same way as that in the Drell-Yan
case in Eq.~(\ref{Gamma-def}). 

Introduce the nuclear broadening of J/$\psi$ transverse momentum
spectrum $\Delta\langle q_T^2\rangle_{{\rm J/}\psi}$, 
\begin{equation}
\Delta\langle q_T^2\rangle_{{\rm J/}\psi}
\equiv \langle q_T^2\rangle^{hA}_{{\rm J/}\psi} 
- \langle q_T^2\rangle^{hN}_{{\rm J/}\psi}\, ,
\label{broad-jp}
\end{equation}
where $\langle q_T^2\rangle^{hN}_{{\rm J/}\psi}$ is the averaged
transverse momentum square of J/$\psi$ production in hadron-nucleon
collisions.  Similar to Eq.~(\ref{alpha-qt}), we derive the nuclear
dependence coefficient $\alpha(A,q_T)$ for J/$\psi$ production in the
small $q_T$ region, 
\begin{eqnarray}
\alpha_{{\rm J/}\psi}(A,q_T) &=& 1+\, 
\frac{1}{\ln(A)} \left[ \ln\left(R^A_{{\rm J/}\psi}\right) 
+ \ln\left(\frac{1}{1+\chi_{{\rm J/}\psi}}\right) \right. 
\nonumber \\ 
&\ & {\hskip 0.6in} \left. 
 + \frac{\chi_{{\rm J/}\psi}}{1+\chi_{{\rm J/}\psi}}\, 
\frac{q_T^2}{\langle q_T^2 \rangle^{hN}_{{\rm J/}\psi} 
               -\Gamma(q_T^L)^{hN}_{{\rm J/}\psi}} 
\right],
\label{a-JPsi}
\end{eqnarray}
where $R^A_{{\rm J/}\psi}\equiv (1/A)\sigma^{hA}_{{\rm J/}\psi}
/\sigma^{hN}_{{\rm J/}\psi}$, and 
$\chi_{{\rm J/}\psi}$ is defined by 
\begin{equation}
\chi_{{\rm J/}\psi} 
\equiv 
\frac{\Delta \langle q_T^2 \rangle_{{\rm J/}\psi} 
     -\Delta\Gamma(q_T^L)_{{\rm J/}\psi}} 
     {\langle q_T^2 \rangle^{hN}_{{\rm J/}\psi} 
     -\Gamma(q_T^L)^{hN}_{{\rm J/}\psi}} 
\approx 
\frac{\Delta \langle q_T^2 \rangle_{{\rm J/}\psi}}
     {\langle q_T^2 \rangle^{hN}_{{\rm J/}\psi}}.
\label{chiJPsi}
\end{equation}
Similar to the Drell-Yan case, $\Gamma(q_T^L)^{hN}_{{\rm J/}\psi}$ and 
$\Delta\Gamma(q_T^L)_{{\rm J/}\psi}$ are perturbatively 
calculable, and much smaller
than $\langle q_T^2 \rangle^{hN}_{{\rm J/}\psi}$ and 
$\Delta\langle q_T^2 \rangle_{{\rm J/}\psi}$,
respectively. 
By comparing Eq.~(\ref{alpha-qt}) with Eq.~(\ref{a-JPsi}), we can see 
that the
$\alpha_{{\rm J/}\psi}(A,q_T)$ for J/$\psi$ production has the same 
universal quadratic dependence on $q_T$:\ 
$\alpha_{{\rm J/}\psi}(A,q_T)=a_2+b_2\,q_T^2$, 
and the parameters $a_2$ and $b_2$ are again completely
fixed by either perturbatively calculable quantities 
(such as $\Delta\langle q_T^2\rangle_{{\rm J/}\psi}$ and 
$\Gamma(q_T^L)_{{\rm J/}\psi}$'s) or independently
measurable quantities (such as the ratio of the total cross section  
$(1/A)\sigma^{hA}_{{\rm J/}\psi}/\sigma^{hN}_{{\rm J/}\psi}$
and the averaged transverse momentum square 
$\langle q_T^2\rangle^{hN}_{{\rm J/}\psi}$). 

For hadron-nucleus collisions at fixed target energies, clear nuclear
suppression for J/$\psi$ total cross sections has been observed
\cite{JPsi-sup,NA50}.  Various theoretical explanations of the
observed J/$\psi$ suppression have been proposed \cite{Dima}.  Unlike
the virtual photon in the Drell-Yan production, the produced $c\bar{c}$
are likely to interact with the nuclear medium before they exit.  If
one assumes each interaction between the $c\bar{c}$ pair
and the nuclear medium is about the {\it same} and can be
treated {\it independently}, one naturally derives the Glauber
formula for the suppression
\begin{equation}
R^A_{{\rm J/}\psi}=
\frac{\sigma_{{\rm J/}\psi}^{hA}/A}{\sigma_{{\rm J/}\psi}^{hN}}
 = {\rm e}^{-\beta L_A}\, , 
\label{JPsiA}
\end{equation}
where $\beta$ is a parameter characterizing the size of suppression,
and $L_A\propto A^{1/3}$ is the effective length of the nuclear
medium.  In terms of Glauber theory of multiple scattering, the
$\beta=\rho_N \sigma_{\rm abs}$ with the nuclear density $\rho_N$ and
an effective absorption cross section $\sigma_{\rm abs}$ for breaking
a J/$\psi$ meson or for changing a coherent pre-resonance $c\bar{c}$
pair to a pair of open charms.  The characteristic feature of the
Glauber formula is a straight line on a semi-log plot of
$\sigma^{hA}_{{\rm J/}\psi}$ vs. $L_A$.  With a large effective
$\sigma_{\rm abs}\sim 6-7$~mb, the 
Glauber formula given in Eq.~(\ref{JPsiA}) fits almost all J/$\psi$
suppression data in hadron-nucleus and nucleus-nucleus collisions
\cite{Dima}, except the strong suppression observed in Pb-Pb
collisions \cite{NA50}. 

If one believes that there is no color-deconfinement or quark-gluon
plasma was formed at present heavy ion collisions at fixed target
energies, one has to overcome at least two obvious difficulties of the 
Glauber formula for the suppression: the size of the effective
absorption cross section $\sigma_{\rm abs}$, and the observed
non-linearity of the suppression on the semi-log plot of
$\sigma^{hA}_{{\rm J/}\psi}$ vs. $L_A$.  
It was argued in Ref.~\cite{BM} that because of the size of the
$c\bar{c}$ pair produced in a hard collision, the J/$\psi$ meson has to
be formed several fermis later after the production of the $c\bar{c}$
pair.  Therefore, the suppression of J/$\psi$ in hadronic collisions
should be a result of multiple scattering between the produced
$c\bar{c}$ pairs and the nuclear medium before the pairs exit medium
\cite{QVZ,BQV}.   Recall the fact that the Glauber formula in
Eq.~(\ref{JPsiA}) is derived for the propagation of a single particle
in a medium.  On the other hand, the pre-resonance J/$\psi$ is a state
of at least two particles (e.g., a $c\bar{c}$ pair), and such a two
particle state can be changed when it goes through the nuclear medium
due to the multiple scattering.  
Just like the random walk for a pair of particles moving
through a medium, multiple scattering increases the relative momentum
between the pairs, and therefore, decreases the effective phase space
for the pairs to form a bound state J/$\psi$ meson.  The observed
suppression should be an immediate consequence of the fact that a
$c\bar{c}$ pair with a {\it larger} relative momentum is more likely
to become a pair of open charm mesons than a bounded J/$\psi$.  In
this picture of J/$\psi$ suppression, the absorption cross section 
$\sigma_{\rm abs}$ in Eq.~(\ref{JPsiA}) is an effective cross section
for breaking a $c\bar{c}$ pair of relative momentum $\Delta
k_{c\bar{c}}$ into a pair of open charm mesons, and clearly, such an
effective cross section depends on the value of $\Delta k_{c\bar{c}}$.
The larger $\Delta k_{c\bar{c}}$ is, the larger the effective
$\sigma_{\rm abs}$ should be.  Because the multiple scattering between
gluons and the pair changes the value of $\Delta k_{c\bar{c}}$ 
while the pair exits the
nuclear medium, the effective $\sigma_{\rm abs}$ cannot be a constant, 
and it should increase as the $c\bar{c}$ passes through the medium.
Therefore, we expect a stronger suppression than that predeicted by
the Glauber formula in Eq.~(\ref{JPsiA}) for a large nuclear medium,
which results into a non-linear dependendence of the suppression on
the semi-log plot \cite{QVZ}.  The quantitative prediction of the
suppression depends on the increase of relative momentum per unit
length for the $c\bar{c}$ pairs in nuclear medium.    
With only two parameters: $\varepsilon^2$, the average relative
momentum square per unit length in medium acquired by the $c\bar{c}$
pair, and $\alpha_F$ which determines the transition probability for a
$c\bar{c}$ pair of relative momentum $\Delta k_{c\bar{c}}$ to form a
bound-state J/$\psi$ meson, it was shown in Ref.~\cite{QVZ} that 
this picture of J/$\psi$ suppression is consistent with all existing
data including the NA50 data on the total J/$\psi$ cross section as
well as its $E_T$ spectrum (or dependence on transverse energy $E_T$
of the J/$\psi$ events).

However, in this paper, we are interested in the general features of
the $\alpha_{{\rm J/}\psi}(A,q_T)$ in hadron-nucleus collisions at
fixed target energies.  Because of the fact that the ratio of
J/$\psi$ total cross sections in {\it hadron-nucleus} collisions at
fixed target energies can be fitted by the Glauber formula in
Eq.~(\ref{JPsiA}), we will not present any detailed model calculation
for the ratio of J/$\psi$ cross sections here \cite{QVZ}.  Instead, 
we will approximate the J/$\psi$ suppression by using the
$R^A_{{\rm J/}\psi}$ defined in Eq.~(\ref{JPsiA}) for our discussions
in the rest of this paper.

Similar to the Drell-Yan process, the nuclear broadening of transverse
momentum square for J/$\psi$ production, 
$\Delta\langle q_T^2\rangle_{{\rm J/}\psi}$, can
be calculated in principle within QCD perturbation theory.  In our
picture of the J/$\psi$ suppresion, the $\Delta\langle
q_T^2\rangle_{{\rm J/}\psi}$ should not be directly tied to the
J/$\psi$ suppression, because the broadening is an effect on the total
momentum of the $c\bar{c}$ pairs while the suppression is a result of
changing the relative momentum of the pairs.  This is why the observed
nuclear broadening of the transverse momentum square for the J/$\psi$
and Drell-Yan production shows a similar $A^{1/3}$ dependence, while
the J/$\psi$ and Drell-Yan total cross sections have a very different
suppression. 

However, due to extra final-state interactions in
J/$\psi$ production and the difference in partonic subprocesses
between the Drell-Yan and J/$\psi$ production,  
$\Delta\langle q_T^2\rangle_{{\rm J/}\psi}$ will depend on 
the gluon-gluon correlation
function in addition to the quark-gluon correlation functions
mentioned in previous sections \cite{LQS2}.  But, for a normal nuclear
target, both these correlation functions should be proportional to the
target length or $A^{1/3}$.  Similar to Eq.~(\ref{dyqt2c}), for the
normal nuclear targets, we can define 
\begin{equation}
\Delta\langle q_T^2\rangle_{{\rm J/}\psi} 
= b_{{\rm J/}\psi}\, A^{1/3}\, ,
\label{JPsi-qt2-a13}
\end{equation}
where $b_{{\rm J/}\psi}$ can be extracted from existing data.  From
Ref.~\cite{JPsi-qt2-exp}, $b_{{\rm J/}\psi} \approx 0.06$~GeV$^2$,
which is larger than $b_{DY}$.  With the averaged transverse momentum
square $\langle q_T^2\rangle^{hN}_{{\rm J/}\psi}\approx
1.68$~GeV$^2$ extracted from the fit in Fig.~\ref{fig4}, 
we have $\chi_{{\rm J/}\psi}$ ranges from 0.08 to 0.22 for most
relevant nuclear targets.  Clearly, $\chi_{{\rm J/}\psi}$ is larger
than $\chi_{DY}$, but, it is still a small number for most targets.

Substituting Eqs.~(\ref{JPsiA}) and (\ref{JPsi-qt2-a13}) into
Eq.~(\ref{a-JPsi}), and taking the small $\chi_{{\rm J/}\psi}$ limit,
we derive the leading contribution to the $\alpha_{{\rm
J/}\psi}(A,q_T)$ for J/$\psi$ production in a normal nuclear medium as 
\begin{equation}
\alpha_{{\rm J/}\psi}(A,q_T) \approx  
1 - \beta\, r_0\,
+ \frac{b_{{\rm J/}\psi}}
       {\langle q_T^2\rangle^{hN}_{{\rm J/}\psi}}
\left[ -1 + \frac{q_T^2}{\langle q_T^2 \rangle^{hN}_{{\rm J/}\psi}} 
\right]  ,
\label{a-JPsi0}
\end{equation}
where $\beta$ was introduced in Eq.~(\ref{JPsiA}) and $r_0$ is a
proportional constant defined as $L_A=r_0\,A^{1/3}$.
In obtaining Eq.~(\ref{a-JPsi0}), we also neglected the very small 
$\Gamma(q_T^L)^{hN}_{{\rm J/}\psi}$.   The leading contribution to 
$\alpha_{{\rm J/}\psi}(A,q_T)$ 
given in Eq.~(\ref{a-JPsi0}) has the same universal quadratic
dependence on $q_T$ as $\alpha_{DY}(A,q_T)$
 in Eq.~(\ref{alpha-qt0}), except a
$q_T$-independent shift in magnitude given by $- \beta
\frac{L_A}{\ln(A)}$.  Since $L_A$ is proportional to $A^{1/3}$ and
$A^{1/3}\approx \ln(A)$, the shift in magnitude is insensitive to the
atomic number $A$.  

Similar to the Drell-Yan case, we obtain from Eq.~(\ref{a-JPsi}), 
\begin{eqnarray}
\alpha_{{\rm J/}\psi}(q_T=0)  &=& 1 
- \frac{\ln(1+\chi_{{\rm J/}\psi})}{\ln(A)} 
+ \frac{1}{\ln(A)} \,
  \ln\left(\frac{\sigma^{hA}_{{\rm J/}\psi}/A}
                {\sigma^{hN}_{{\rm J/}\psi}}\right)
\nonumber \\
&\approx &
1- \frac{b_{{\rm J/}\psi}}
        {\langle q_T^2 \rangle^{hN}_{{\rm J/}\psi}}
 - \beta\, r_0 \, ,
\label{a-Jpsi-qt0}
\end{eqnarray}
and it shows that the $\alpha_{{\rm J/}\psi}(q_T=0)$ is insensitive 
to the atomic number $A$ for normal nuclear targets.  Corresponding to 
$\alpha_{{\rm J/}\psi}(q_T^c)=1$, the
critical value $q_T^c$ for J/$\psi$ production is given by
\begin{eqnarray}
(q_T^c)^2 &=& \langle q_T^2 \rangle^{hN}_{{\rm J/}\psi}\left( 
            \frac{1+\chi_{{\rm J/}\psi}}{\chi_{{\rm J/}\psi}}\right)
\left[\ln(1+\chi_{{\rm J/}\psi}) 
- \ln\left(\frac{\sigma^{hA}_{{\rm J/}\psi}/A}
                {\sigma^{hN}_{{\rm J/}\psi}}\right) \right]
\nonumber \\
&\approx &
\langle q_T^2 \rangle^{hN}_{{\rm J/}\psi}
\left[ 1 + \frac{\beta\, r_0}{b_{{\rm J/}\psi}}\,
           \langle q_T^2 \rangle^{hN}_{{\rm J/}\psi} \right]\, ,
\label{qtc-Jpsi}
\end{eqnarray}
which is again insensitive to the atomic number $A$.

\begin{figure}
\begin{minipage}[t]{2.0in}
\epsfig{figure=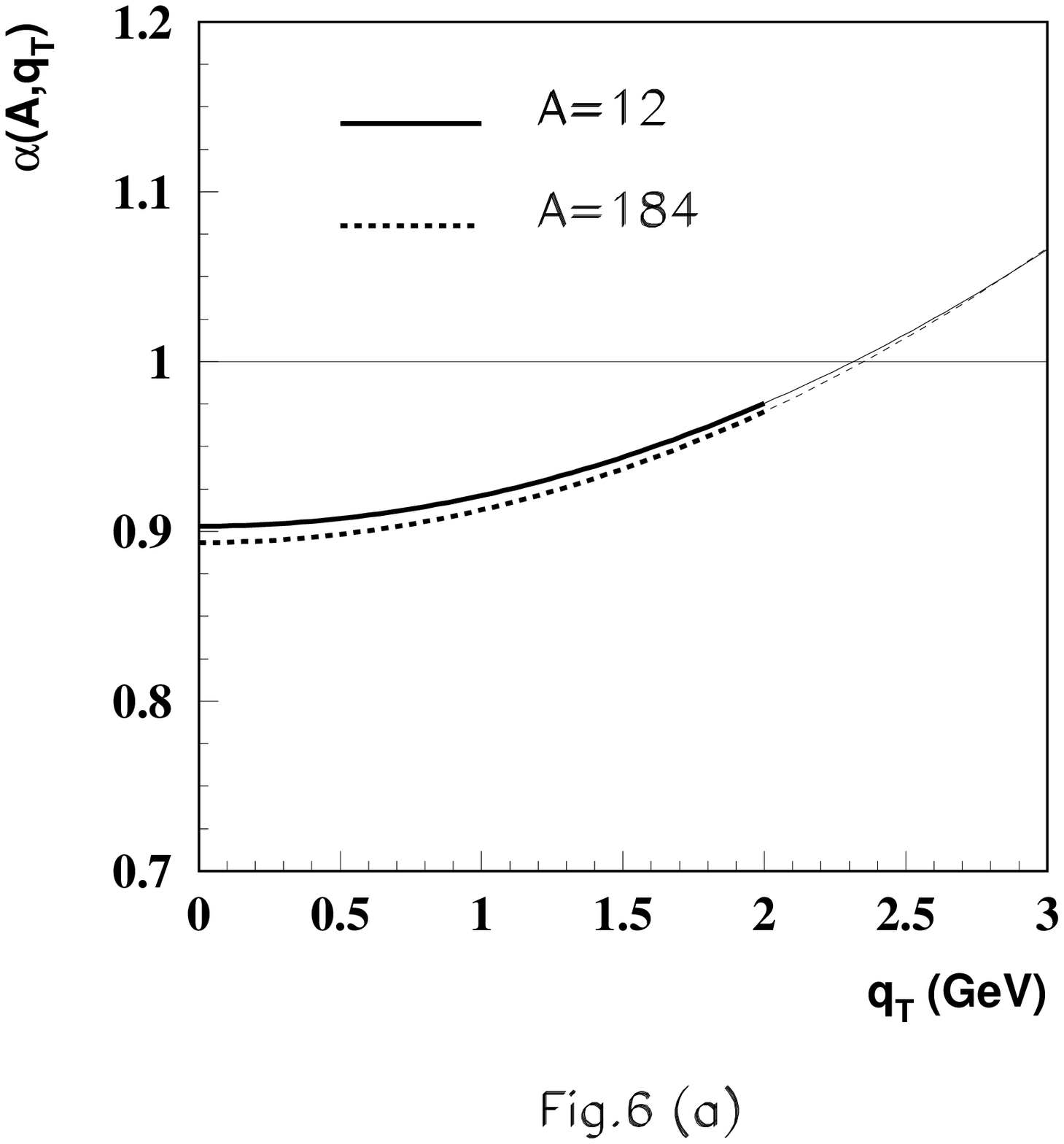,width=1.0in}
\end{minipage}
\hfill
\begin{minipage}[t]{2.0in}
\end{minipage}
\epsfig{figure=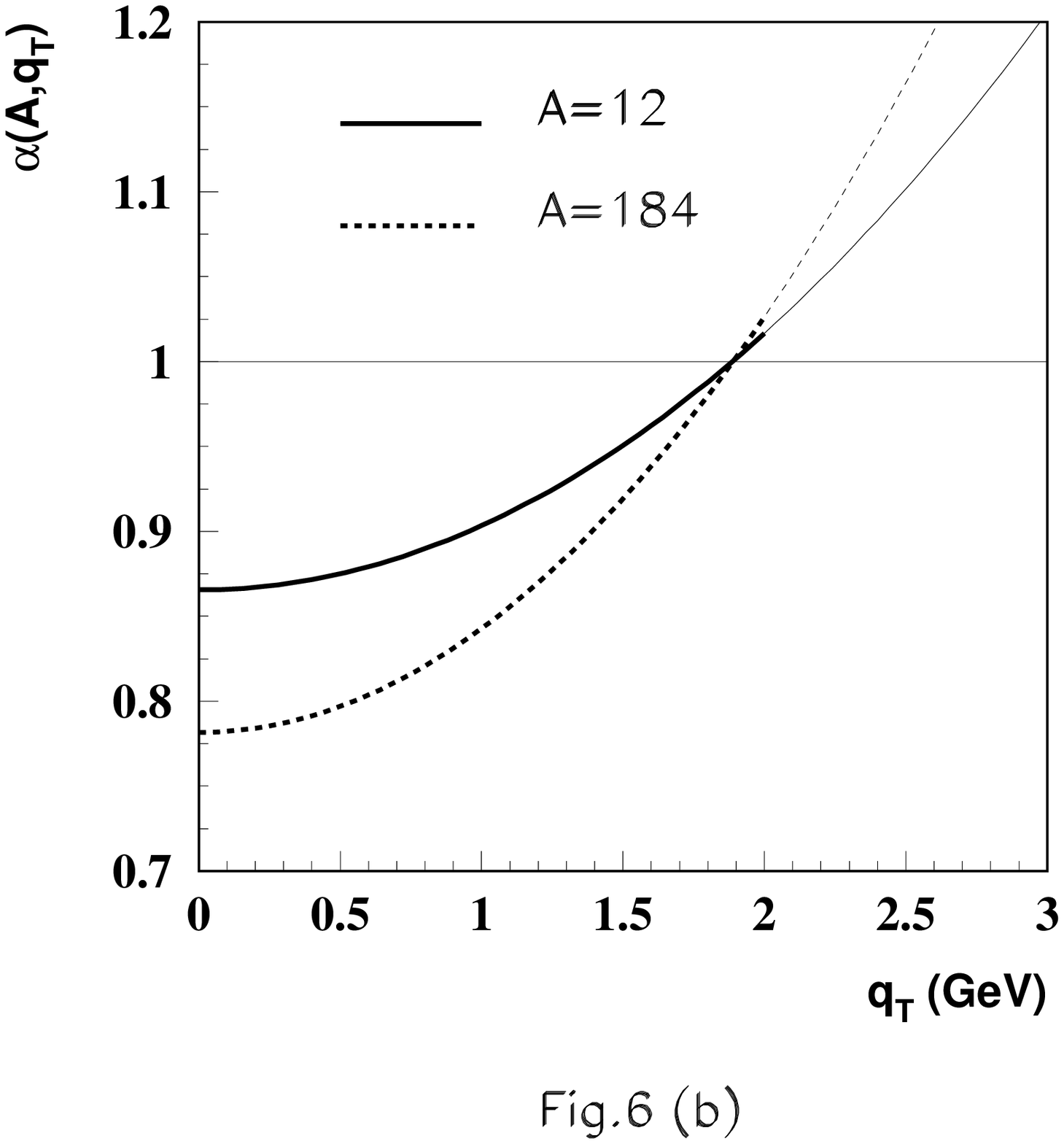,width=1.0in}
\caption{$\alpha(A,q_T)$ for J/$\psi$ production, defined in
Eq.~(\protect\ref{a-JPsi}), as a function of $q_T$ for normal color
confined nuclear targets (a), and for possible color deconfined
nuclear medium (b).} 
\label{fig6}
\end{figure}

In Fig.~\ref{fig6}, we plot the $\alpha_{{\rm J/}\psi}(q_T)$, defined
in Eq.~(\ref{a-JPsi}), as a function of $q_T$ for different nuclear
targets.  In plotting Fig.~\ref{fig6}a, we used $\langle q_T^2
\rangle^{hN}_{{\rm J/}\psi}=1.68$~GeV$^2$, and 
$\Delta\langle q_T^2 \rangle_{{\rm J/}\psi} =
0.06 A^{1/3}$~GeV$^2$.  The solid and dashed lines correspond to the nuclear 
target C
($A$=12) and W ($A$=184), respectively.  As expected from
Eq.~(\ref{a-JPsi0}), Fig.~\ref{fig6}a demonstrates that 
$\alpha_{{\rm J/}\psi}(A,q_T)$ for J/$\psi$ production has the 
same universal functional form as $\alpha_{DY}(A,q_T)$, and it also 
has scaling on $A$.  

In Fig.~\ref{fig6}a, the thin lines correspond to the large $q_T$ region
($q_T\ge 2$~GeV), where we expect that perturbative
calculations of the nuclear dependence should be valid.  However, such
calculations are not available yet.

Since the suppression of J/$\psi$ total cross sections in
hadron-nucleus collisions corresponds to
a $q_T$-independent shift in the value of 
$\alpha_{{\rm J/}\psi}(A,q_T)$, as we explained earlier, 
the potential nuclear dependence for the shape of $\alpha_{{\rm
J/}\psi}(A,q_T)$ is most sensitive to the nuclear broadening
$\Delta\langle q_T^2\rangle_{{\rm J/}\psi}$. 
Although QCD calculation of the nuclear broadening of J/$\psi$
transverse momentum square is not available yet, similar to the
Drell-Yan case, we expect the broadening $\Delta\langle
q_T^2\rangle_{{\rm J/}\psi}$ to be proportional to the
``soft-hard'' parton-parton correlation functions.  Different from the
``double-hard'' parton-parton correlation functions, both position
variables for soft gluons, such as the $dy_1^-$ and $dy_2^-$ in 
Eq.~(\ref{dyTq}), have no exponential oscillation factors.  Therefore,
for a color deconfined nuclear medium, both position variables can be
as large as the size of the medium; and in principle, the nuclear
broadening for J/$\psi$ production can be enhanced as much as
$A^{2/3}$, or $\Delta\langle q_T^2\rangle_{{\rm J/}\psi} \approx 
b \, A^{2/3}$, if we let the color correlation length be proportional 
to $A^{1/3}$.  Notice, the proportional parameter $b$ given here is
not necessary the same as the $b_{{\rm J/}\psi}$ extracted from data
on a normal nuclear medium.  Since we are mainly interested in the
$A$-dependence of the $\alpha(A,q_T)$, we will approximate 
the $b= b_{{\rm J/}\psi}$ for following discussions.

In a color deconfined medium, the $\chi_{{\rm
J/}\psi}$ in Eq.~(\ref{a-JPsi}) is no longer a small number, and its  
$A^{2/3}$-dependence can not be canceled by the $\ln(A)$.
Consequently, the $\alpha_{{\rm J/}\psi}(A,q_T)$ should become very
sensitive to the atomic weight $A$ (or the medium size).
In Fig.~\ref{fig6}b, we plotted the $\alpha_{{\rm J/}\psi}(A,q_T)$ 
with the same parameters as those used to plot Fig.~\ref{fig6}a,
except $\Delta\langle q_T^2 \rangle_{{\rm J/}\psi} = 
0.06 A^{2/3}$~GeV$^2$, which mimics a color deconfined nuclear medium.
Notice that in 
plotting Fig.~\ref{fig6}b, we did not include a possible change in
$A$-dependence of the ratio of J/$\psi$ total cross sections, since it 
contributes only to a $q_T$-independent shift in magnitude.
From Fig.~\ref{fig6}, it is clear that $A$-dependence of the 
$\alpha_{{\rm J/}\psi}(q_T)$ for J/$\psi$ production can be a
sensitive probe for the color deconfinement of the nuclear medium.

\section{Discussions and Summary}
\label{sec:5}

In this section, we discuss the predicting power of the nuclear
dependence coefficient $\alpha(A,q_T)$ for the Drell-Yan and J/$\psi$
production derived in the previous two sections, and also discuss 
the uncertainties in the analytical
expressions.  Finally, we summarize our main conclusions.

The predicting power of the nuclear dependence coefficient 
$\alpha(A,q_T)$ in Eqs.~(\ref{alpha-qt}) and
(\ref{a-JPsi}) is its universal quadratic dependence on $q_T$, and the
fact that all parameters are completely fixed by 
either calculable or
independently measurable quantities: (1) the ratio of {\it total}
cross sections, (2) the averaged transverse momentum square in
hadron-{\it nucleon} collisions, (3) the nuclear broadening of the
averaged transverse momentum square, and (4) the perturbatively
calculable $\Gamma(q_T^L)$'s.  The ratio of {\it total}
cross sections, $\sigma^{hA}/\sigma^{hN}$, can be either measured 
in experiments or calculated in a theory.  The ratio is independent 
of $q_T$, and therefore, it does not affect the shape of $\alpha(A,q_T)$. 
On the other hand, the averaged transverse momentum square in
hadron-{\it nucleon} collisions, $\langle q_T^2 
\rangle^{hN}$, does not have any nuclear dependence and can be
independently measured.  The nuclear broadening of the averaged transverse 
momentum square, $\Delta\langle q_T^2 \rangle$, can be factorized
into multi-parton correlation functions and perturbatively calculable
partonic parts.  These non-perturbative multi-parton correlation
functions are universal and can be extracted from the nuclear
dependence in the large $q_T$ region.  Therefore, adding the small
corrections from the perturbatively calculable $\Gamma(q_T^L)$'s,
we can actually predict the $\alpha(A,q_T)$ for the Drell-Yan and
J/$\psi$ production.  
Fig.~\ref{fig4}b is an example of our prediction and its  
comparison with E772 data. 

The main uncertainty in our predictions for $\alpha(A,q_T)$ comes from 
the size of the QCD resummation for potential large logarithms, such as  
$(\alpha_s\ln^2(Q^2/q_T^2))^n$.  For the Drell-Yan production at both
collider and fixed target energies, our formula for the
$\alpha_{DY}(A,q_T)$ should be valid or at least a good approximation,
as long as $Q^2$ is not too large and $q_T\leq 2.5$~GeV.  For J/$\psi$
production, because $M_{{\rm J/}\psi}$ is relatively small in
comparison with the $Q$ for a typical Drell-Yan pair, 
our formula for $\alpha_{{\rm J/}\psi}(A,q_T)$ should be
valid for $q_T\leq 2$~GeV.  When $q_T$ is larger, 
$\alpha_{{\rm J/}\psi}(A,q_T)$ should be calculable in 
perturbative QCD.

Another possible uncertainty in our predictions is the asymptotic
value of normal parton distributions as $x\rightarrow 0$ at low
energy, as seen in Eq.~(\ref{c-lambda}).  The precise asymptotic value
of the parton distributions as $x\rightarrow 0$ itself is a very
interesting quantity, which can provide rich information on parton 
saturation \cite{Mueller}. For the quantities we discussed in this
paper, different asymptotic values
of the parton distributions result into different normalizations of the
``double-hard'' parton-parton correlation functions.  Since the
contributions from the ``double-hard'' correlation functions to the
$\alpha_{DY}(A,q_T)$ in the large $q_T$ region is as important as those
from  the ``soft-hard'' correlation functions \cite{Guo1},
different asymptotic values of the parton distributions can result
into a different crossing point between the low $q_T$ spectrum and the
high $q_T$ spectrum, and consequently, a slightly different value of
the $q_T^L$.  For example, a smaller value 
of $xG(x)|_{x\sim 0}$ corresponds to a slightly larger value of
$q_T^L$, which actually describes the data slightly better.  Although 
current data are still not sufficient to provide the precise values of 
parton distributions as $x\rightarrow 0$, such asymptotic values can
be measured in principle, and corresponding uncertainties can be
fixed.  

The four key quantities for determining the $\alpha(A,q_T)$: 
$\sigma^{hA}/\sigma^{hN}$, $\langle q_T^2 \rangle^{hN}$, 
$\Delta\langle q_T^2 \rangle$, and $\Gamma(q_T^L)$'s 
can depend on beam energy, $P_{\rm beam}$,  and $x_F$, and also 
the value of $Q$ in the case of the Drell-Yan production.  For
example, the dependence of $\alpha(A,q_T)$ on the beam energy can be 
qualitatively understood as follows.  For given values of $x_F$ (and 
$Q$ in the case of the Drell-Yan production), a larger 
$P_{\rm beam}$ (or $\sqrt{S}$) leads
to a larger $\langle q_T^2\rangle^{hN}$ because of the larger phase space.
From Eq.~(\ref{alpha0}) [or Eq.~(\ref{a-Jpsi-qt0})], we conclude that
when $P_{\rm beam}$ increases, the intersection, $\alpha(q_T=0)$,
with the vertical axis at $q_T=0$ in Fig.~\ref{fig4} [or
Fig.~\ref{fig6}a] moves closer to one.  On the other hand,
from Eq.~(\ref{qtc}) [or Eq.~(\ref{qtc-Jpsi})], the increase of the
beam energy leads to a larger value of the critical point $q_T^c$ 
at which the $\alpha(A,q_T^c)=1$.  Therefore, the curve of 
$\alpha(A,q_T)$ becomes flatter when $P_{\rm beam}$ increases, which
is consistent with the data from NA10 \cite{NA10-pi}.
Similar qualitative conclusions can be derived for the
$x_F$-dependence (or $Q$-dependence in the case of 
the Drell-Yan production) of $\alpha(A,q_T)$.  In any case, 
one should take into account the
possible $x_F$ dependence of all four key quantities that determine
$\alpha(A,q_T)$.  

\begin{figure}
\epsfig{figure=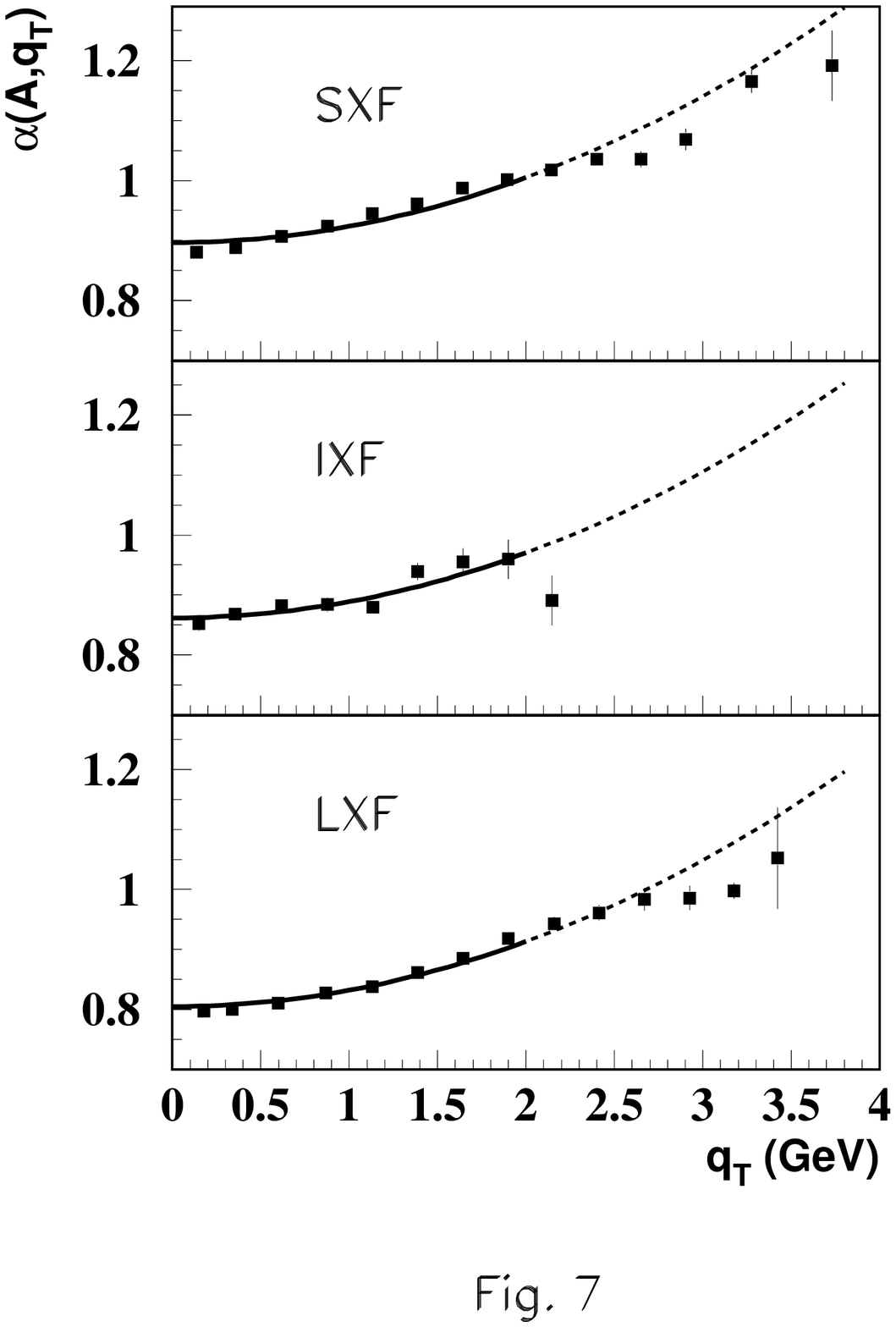,width=1.1in}
\caption{Comparison of fits, using our universal quadratic form for
$\alpha(A,q_T)$ in Eq.~(\protect\ref{a-JPsi}), with J/$\psi$ data from 
Fermilab E866 collaboration \protect\cite{E866} in three regions of 
different $x_F$.}
\label{fig7}
\end{figure}

Recent data from Fermilab experiment E866 shows that for different
regions of $x_F$, the $\alpha_{{\rm J/}\psi}(A,q_T)$ for J/$\psi$
production have similar shapes in $q_T$-dependence, but, different
magnitudes \cite{E866}.  It was found that for three $x_F$ regions
(small, intermediate, and large), the magnitude of 
$\alpha_{{\rm J/}\psi}(A,q_T)$ decreases as $x_F$ increases.  
This phenomenon is  consistent with our prediction
of  $\alpha_{{\rm J/}\psi}(A,q_T)$ in Eq.~(\ref{a-JPsi}) [or
Eq.~(\ref{a-JPsi0})].  According to Eq.~(\ref{a-JPsi}), the
suppression in J/$\psi$ {\it total} cross sections represents a
$q_T$-independent shift in the magnitude of 
$\alpha_{{\rm J/}\psi}(A,q_T)$.  The larger the suppression, the
smaller $\alpha_{{\rm J/}\psi}(A,q_T)$.
Because of limited $x_F$ range, the ratio of total cross sections,
$R^A_{{\rm J/}\psi}\equiv 
(1/A)\sigma^{hA}_{{\rm J/}\psi}/\sigma^{hN}_{{\rm J/}\psi}$ in 
Eq.~(\ref{a-JPsi}), becomes $x_F$ dependent.  It is an 
experimental fact that the larger 
$x_F$ is, the more suppression for J/$\psi$ production 
(or smaller $R^A_{{\rm J/}\psi}$) \cite{Jpsi-xf-exp}.  
Therefore, from Eq.~(\ref{a-JPsi}), the larger $x_F$, the smaller  
$\alpha_{{\rm J/}\psi}(A,q_T)$, which is consistent with experimental 
data \cite{E866}.  Although we do not have the four key quantities for
different $x_F$ regions to make absolute predictions, we can
still test the universality (or consistency) of the 
$\alpha_{{\rm J/}\psi}(A,q_T)$ by using the quadratic form in $q_T$ 
shown in Eq.~(\ref{a-JPsi}) to fit the data.  In Fig.~\ref{fig7}, we plot
E866 data on $\alpha_{{\rm J/}\psi}(A,q_T)$ in three separate
$x_F$ regions: small (SXF), intermediate (IXF), and large (LXF).  We
also plot in Fig.~\ref{fig7} our fits to these data with a
universal quadratic form in $q_T$, as shown in Eq.~(\ref{a-JPsi}).  It
is clear from Fig.~\ref{fig7} that our universal function for 
$\alpha_{{\rm J/}\psi}(q_T)$ is consistent with all data in the small
$q_T$ region, which covers $q_T < q_T^L \sim M_{{\rm J/}\psi}/2$.  
Also, as expected,
our formula for $\alpha_{{\rm J/}\psi}(A,q_T)$ deviates from the data
when $q_T> M_{{\rm J/}\psi}/2$.  Like the Drell-Yan case, shown in
Fig.~\ref{fig4}, it will be very interesting to calculate the nuclear
dependence of the J/$\psi$ transverse momentum spectrum in the large $q_T$
region to test the QCD dynamics.

In summary, we derived an analytic formula for the nuclear
dependence coefficient $\alpha(A,q_T)$ for both the Drell-Yan and J/$\psi$
production in the small $q_T$ region.  The formula has a universal
quadratic dependence on $q_T$, and all parameters are completely
determined by four key quantities: $\sigma^{hA}/\sigma^{hN}$,
$\langle q_T^2 \rangle^{hN}$, $\Delta\langle q_T^2 \rangle$, and
$\Gamma(q_T^L)$'s.  These quantities can be either calculated in QCD
perturbation theory or independently measured in other experiments.  
We explicitly demonstrated that our $\alpha(A,q_T)$ is
consistent with existing data.

Furthermore, we showed that our $\alpha(A,q_T)$ is extremely 
{\it insensitive} to the atomic weight $A$ for a normal nuclear
target.  However, for a color deconfined nuclear medium, the
$\alpha(A,q_T)$ becomes strongly dependent on $A$ (or the medium 
size). Therefore, the 
$A$-dependence of $\alpha(A,q_T)$ can be a potential observable 
for detecting a color deconfined system.  
 
\section*{Acknowledgment}
We thank M.J. Leitch, J.M. Moss, and J.-C. Peng for helpful 
communications about experiments and data. This work was supported in
part by the U.S. Department of Energy under Grant Nos.
DE-FG02-87ER40731 and  DE-FG02-96ER40989.


\begin{references}
\bibitem{Cronin}
J.W. Cronin {\it et al}, Phys.Rev. D {\bf 11}, 3105(1975).

\bibitem{Cronin-data}
D. Antreasyan, {\it et al}, Phys. Rev. {\bf D 19}, 764 (1979).
P.B. Straub,  {\it et al}, Phys. Rev. Lett. {\bf 68}, 452 (1992).

\bibitem{NA10}
 P. Bordalo, {\it et al}, NA10 Collaboration, 
 Phys. Lett. {\bf B193}, 373(1987).

\bibitem{E772}
 D. M. Alde, {\it et al}, E772  Collaboration, 
 Phys. Rev. Lett. {\bf 66}, 2285 (1991);
 Phys. Rev. Lett. {\bf 64}, 2479 (1990).

\bibitem{Eloss}
 M.A. Vasiliev {\it et al}, E866 Collaboration,  
 Phys. Rev. Lett. 83, 2304(1999).

\bibitem{E866}
 M.J. Leitch {\it et al}, E866 Collaboration,   nucl-ex/9909007.

\bibitem{Jpsi-Et}
 M. C. Abreu {\it et al}, NA38  Collaboration,  
 Phys. Lett. {\bf B368}, 230 (1996);
 C. Baglin {\it et al},  NA38  Collaboration, Phys Lett. {\bf B 251},
 465 (1990) and  Phys Lett. {\bf B 368}, 230 (1996).

\bibitem{NA3}
 J. Badier {\it et al}, NA3 Collaboration,  
 Z. Phys. {\bf C 20}, 101  (1983).

\bibitem{alpha-thy}
 S. Gavin and M. Gyulassy, Phys. Lett. {\bf B214}, 241 (1988);
 J. H\"{u}fner, Y. Kurihara, and H.J. Pirner, Phys. Lett. {\bf B215}, 
 218 (1988);
 J.P. Blaizot and J.-Y. Ollitrault, Phys. Lett. {\bf B217}, 386, 
 392 (1989);
 D. Kharzeev, M. Nardi, and H. Satz, Phys. Lett. {\bf B405}, 14 (1997).

\bibitem{LQS1}
 M. Luo, J.-W. Qiu, and G. Sterman, Phys. Lett. {\bf B279}, 377 (1992);
 M. Luo, J.-W. Qiu, and G. Sterman, Phys. Rev. D{\bf 50}, 1951 (1994).

\bibitem{Guo1}
 X.-F. Guo, Phys. Rev. D{\bf 58}, 036001 (1998).

\bibitem{BMuller}
R.J. Fries, B. M\"{u}ller, A. Sch\"{a}fer, and E. Stein,
Phys. Rev. Lett. {\bf 83}, 4261(1999). 

\bibitem{GQZ1}
 X.-F. Guo, J.-W. Qiu, and X.-F. Zhang, hep-ph/9911476.

\bibitem{CSS-fac}
 J.C. Collins, D.E. Soper, and G. Sterman, in ``Perturbative Quantum
Chromodynamics'', ed. A.H. Mueller (World Scientific, 1989).

\bibitem{QS-fac}
J.-W. Qiu and G. Sterman, Nucl. Phys. {\bf B353}, 137 (1991).

\bibitem{LQS2}
 M. Luo, J.-W. Qiu, and G. Sterman, Phys. Rev. D{\bf 49}, 4493 (1994). 

\bibitem{Satz}
T. Matsui and H. Satz, Phys. Lett. {\bf B178},416 (1986).

\bibitem{JPsi-sup}
M. C. Abreu {\it et al}, NA38  Collaboration, Phys. Lett 
{\bf B 444}, 516 (1998);
M. C. Abreu {\it et al}, NA38  Collaboration, Phys. Lett 
{\bf B 449}, 128 (1999);
M. C. Abreu {\it et al}, NA51  Collaboration,   
Phys. Lett. {B438}  35 (1998).

\bibitem{NA50}
M. C. Abreu {\it et al}, NA50  Collaboration, Phys. Lett 
{\bf B 410}, 337 (1997);
M. C. Abreu {\it et al}, NA50  Collaboration,   
Phys. Lett. {B450}  456 (1999).

\bibitem{Guo2} 
 X.-F. Guo, Phys. Rev. D {\bf 58}, 114033 (1998).

\bibitem{ESW} 
 R.K. Ellis, J. Sterling,  and B. Webber, `` QCD and Collider Physics '',
(Cambridge, U.K. 1996 ).

\bibitem{Resum} 
G. Atarelli, G. Parisi, and R. Petronzio, 
 Phys. Lett. {\bf B76}, 351 (1978);
F. Khalafi and W.J. Stirling, 
 Z. Phys. {\bf C18}, 315 (1983); 
M. Beneke and V. M. Braun, Nucl.Phys. {\bf B454}, 253 (1995). 

\bibitem{CSS-resum}
J.C. Collins, D.E. Soper and G. Sterman, 
 Nucl. Phys. {\bf B308}, 833 (1988).

\bibitem{Davies}
C.T.H. Davies and W.J. Stirling, Nucl. Phys. {\bf B244},
337 (1984);
C.T.H. Davies, W.J. Stirling, and B.R. Webber, Nucl. Phys. {\bf B256},
413 (1985).

\bibitem{Yuan}
G.A. Ladinsky and C.P. Yuan, Phys. Rev. {\bf D50}, R4239 (1994).

\bibitem{DY-800}
P. L.McGaughey {\it et al}, E772 Collaboration, Phys. Rev. {\bf D 50}, 
3038 (1994).

\bibitem{CFS} 
 A. S. Ito {\it et al}, CFS Collaboration, 
 Phys. Rev.  {\bf D 23}, 604 (1981). 

\bibitem{Berger}
 E.L. Berger, L. Gordon, and M. Klasen, Phys. Rev. {\bf D58}, 074012
(1998) and hep-ph/9909446. 

\bibitem{DY-LPT}
P. Arnold, R.K. Ellis  and  M.H. Reno  Phys. Rev. {\bf D40},  912, (1989),
and erratum Nucl. Phys. {\bf B330}, 284 (1990);
R. J. Gonsalves, J. Pawlowski, and C. F. Wai, Phys. ReV. {\bf D40}, 225
(1989). 

\bibitem{Guo3}
 X.-F. Guo, X.-F. Zhang, and W. Zhu, Phys. Lett. {\bf B476}, 316(2000). 

\bibitem{MQ}
A.H. Mueller and J.-W. Qiu, Nucl. Phys. {\bf B268}, 427(1986).

\bibitem{xG}
R. Baier, Yu. L. Dokshitzer, A.H. Mueller, S. Peign\'{e} and D.
Schiff, Nucl. Phys. {\bf B484}, 265 (1997).

\bibitem{Mueller}
A.H. Mueller, Columbia preprint CU-TP-937, hep-ph/9904404.

\bibitem{E772-moss}
J. M. Moss and  P. L. McGaughey, private communications.

\bibitem{E771}
T. Alexopoulos {\it et al}, E771 Collaboration, Phys. Rev. {\bf D 55}
3927 (1999).  

\bibitem{Dima}
 D. Kharzeev, Nucl.Phys. {\bf A638}, 279c(1998) and the references 
 therein; S. Raha and B. Sinha, {\bf B198}, 543(1987); {\bf  B218}, 
413(1989).   

\bibitem{BM}
S.J. Brodsky and A.H. Mueller, Phys. Lett. B{\bf 206}, 685 (1988).

\bibitem{QVZ}
 J.-W. Qiu, J.P. Vary, and X.-F. Zhang, in the proceedings of the APS
Centennial Meeting, Atlanta, GA, March 1999 (AIP Press, to appear);
hep-ph/9809442.

\bibitem{BQV}
 C. Benesh, J.-W. Qiu, and J.P. Vary, Phys. Rev. C {\bf 50}, 1015
 (1994). 

\bibitem{JPsi-qt2-exp}
 P. L. McGaughey, J. M. Moss, and J. C. Peng, hep-ph/9905409.
 and reference therein.

\bibitem{NA10-pi}
 P. Bordalo {\it et al.},  NA10 Collaboration, 
 Phys. Lett. {\bf B193}, 373  (1987).

\bibitem{Jpsi-xf-exp}
 D. M. Alde {\it et al}, E772 Collaboration,  
   Phys. Rev. Lett {\bf 66}, 133 (1991); 
 M. S. Kowitt {\it et al}, E789 Collaboration,  
   Phys. Rev. Lett. 72, 1318 (1994), and 
   Phys. Rev. {\bf D52}, 1307 (1995).

\end{references}
\end{document}